\documentclass[11pt]{article}
\pdfoutput=1
\usepackage{ifpdf}
\usepackage{jcappub} 
\usepackage{tikz}
\usepackage[T1]{fontenc} 
\usepackage{tabularx}
\usepackage{natbib}
\usepackage{ae,aecompl,bm}
\usepackage{xcolor}

\usetikzlibrary{positioning}
\usepackage{subcaption}
\newcommand{\be}{\begin{equation}}
\newcommand{\ee}{\end{equation}}
\newcommand{\F}{\mathcal{F}} 

\def\n{{\bf{\hat{n}}}}
\def\k{{\bf{\hat{k}}}}
\def\A{\mathcal{A}}
\def\vtheta{\vec {\mathbf{\theta}}}
\def\vu{\vec{\bf{U}}}
\def\kperp{{\bf{k_{\perp}}}}
\def\kpar{{k_{\parallel}}}
\title{Constraining dark energy using the cross correlations of weak lensing with post-reionization probes of neutral hydrogen }

\author[a]{Chandrachud B.  V. Dash, }
\author[a, 1]{Tapomoy Guha Sarkar \note{Corresponding author.}}

\affiliation[a]{Department of Physics, Birla Institute of Technology and Science, Pilani, 333031, \\Rajasthan, India.}

\emailAdd{cb.vaswar@gmail.com}
\emailAdd{tapomoy@pilani.bits-pilani.ac.in}

\abstract{We investigate the prospects of detecting the cross correlation of  CMBR weak-lensing convergence field with the large scale tracers of the underlying dark matter distribution in the post-reionization epoch. The cross-correlation is then used to make error projections for dark energy equation of state (EoS)for models with a time evolving dark energy.
We study the cross-correlation angular power spectrum of the weak-lensing field with the Lyman-$\alpha$ forest and the redshifted HI 21 cm signal from the post reionization epoch. The angular power spectra is expressed as a line of sight average over the tomographic slices. We find that on  using multiple  $400$ hrs observation with an extended  uGMRT like instrument or with a BOSS like survey with  quasar (QSO) density of $16 {\rm deg}^{-2}$  the  cross-correlation with weak-lensing convergence field covering half the sky can be detected at a very high SNR ($>20$).
The cross-correlation of  weak-lensing with Lyman-$\alpha$ forest allows the $1-\sigma$ errors on the dark energy EoS parameters for different parametrizations to be constrained at a level of precision comparable to combined Planck+SNIa+BAO+HST projections.
The 21-cm weak-lensing cross-correlation is found to provide strong constraints on the present value of the dark energy EoS parameter at $4\%$
for the 7CPL model.  The constraints on $w_a$ is comparable ($\sim 12\%$) for models other than the 7CPL model. We also find that the CPL parametrization may not be the best constrained parametrization for dark energy evolution.
The cross-correlation of CMBR weak-lensing with the  post-reionization probes of neutral hydrogen thus holds the potential to give us valuable understanding about the nature of evolving dark energy.}

\begin{document}
\maketitle
\flushbottom

\section{Introduction}
A host of independent observations in recent times have indicated that  the  expansion of the Universe is accelerating \citep{amendola_tsujikawa_2010, durrer2008dark}. 
The underlying cause of what is driving this late time acceleration is, however, still an open question. The explanation in the  standard paradigm of cosmology assumes  a dark energy component with an equation of state (EoS)  $ ~p/\rho = w (< -1/3)$. 
Precision cosmological measurements indicate that the Universe contains approximately $\sim 70 \%$ of the energy density in the form of dark energy \citep{Perlmutter_1997, Spergel_2003, Hinshaw_2003, Ata_2018} and the remaining $\sim 30 \%$ in the form of non-relativistic matter (both baryonic matter and dark matter). A natural  candidate for constant dark energy is the cosmological constant $\Lambda$.  This model with  $w= -1$ is well tested by many observations.  In this model the cosmological constant $\Lambda$ is to be interpreted as a non-zero vacuum energy density  \citep{Padmanabhan_2003}.
While there are theoretical difficulties pertaining to the cosmological constant (like the `fine tuning' problem), recent results from low redshift measurements of $H_0$ \citep{Riess_2016} also contradict  the Planck-2015 predictions for flat $LCDM$ model.
Further, there are indications that a  varying dark energy model maybe  preferable over the concordance $LCDM$ model \citep{Zhao_2017} at a high level of statistical significance. Our understanding of the cosmic acceleration thus still remains cloaked in mystery.
Dark energy models differing from the standard cosmological constant typically involves a scalar field \citep{Ratra-Peebles_1988, Turner-White_1997, Steinhardt_1998, Picon-Mukhanov_2001, Bento-Sen_2002, TRC-paddy_2002, SenSen_2001, Paddy-Bagla_2003, Mata_1999, Sami_2007, Kujat_2006, Gannouji_2010, Caldwell_2005} whose dynamics with suitable initial conditions  is used to model  the cosmic acceleration. 
It is generally a difficult program to constrain the immense diversity of such scalar field models from observations. It is convenient to  use some parametrization of these models which mimic their general behaviour. The dynamic EoS with $ p/\rho = w (z)$ is one such commonly used parametrization \citep{PhysRevD.65.123003, Thakur_2012, PhysRevD.93.103503,  CHEVALLIER_2001, Paddy-Bagla_2003}.
 The most popular and widely used parametrization of the EoS is  a two-parameter model by  Chavallier-Linder-Polarski (CPL), \citep{CHEVALLIER_2001, PhysRevLett.90.091301}. In this study we shall also use two important variants of the CPL model called the 7CPL model \citep{PhysRevD.93.103503} and Brboza-Alcaniz (BA) model \citep{Barboza_2008}. 
 
Weak gravitational lensing by intervening large scale
structure \citep{Bartelmann_2001, waerbeke2003, MUNSHI_2008} distorts the images of distant background sources,  over large angular scales. This is  caused by the deflection of light by the fluctuating gravitational field created by the intervening overdensity field. Precise quantitative measurement of these distortions opens a window towards our understanding of the large scale matter distribution and geometry of the Universe. Late time cosmic history is governed largely by dark energy, either through 
a modification of the growing mode of density perturbations or through clustering properties of dark energy or both. Weak-lensing studies can be used to impose constraints on dark energy models \cite{knox2004weak, Hoekstra_2008} since the lensing distortions manifest as a line of sight integral of a `kernel' which is sensitive to background evolution and structure formation.
Weak-lensing also distorts the CMBR photon distribution and manifests as secondary anisotropy in the CMBR maps \citep{LEWIS_2006}.
The CMBR temperature and polarization maps may be used to extract the effect of lensing \citep{Hu_2001, Seljak_1999, Hu_2002}. Delensing of the CMBR is crucial in quantifying the imprint of gravitational waves in the B modes \cite{Kamionkowski_2016}.

Neutral hydrogen (HI) in the post-reionization epoch ($z < 6$) \citep{poreion0, poreion1, poreion2, poreion3,  poreion4} is housed in two important astrophysical systems of interest.
The bulk of the neutral gas is found  in  the dense self shielded Damped Lyman-$\alpha$ (DLA) systems \citep{wolfe05, proch05}.
These DLA clouds are the source of the redshifted 21-cm signal, to be seen in emission.
Intensity mapping of the large scale HI distribution using observations the  redshifted
21-cm radiation \citep{Furlanetto_2006, Pritchard_2012} aims to map out the collective diffuse emission without resolving the individual DLA sources \cite{Bull_2015}.
The statistics of these intensity maps is a potentially rich probe of cosmological background evolution and large scale structure formation \citep{param1, param2, param3, param4}.
Several studies look at the possibility of using  21-cm intensity mapping  to constrain dark energy models \citep{Hussain_2016, dinda2018}.
It is also a key science goal of many radio telescopes like the GMRT \footnote{http://gmrt.ncra.tifr.res.in/} OWFA\footnote{https://arxiv.org/abs/1703.00621}, MEERKAT\footnote{http://www.ska.ac.za/meerkat/},  MWA\footnote{https://www.mwatelescope.org/}, CHIME\footnote{http://chime.phas.ubc.ca/}, and SKA\footnote{https://www.skatelescope.org/}  to detect the cosmological 21-cm signal for a tomographic imaging \cite{Mao_2008} at observing frequencies $ \nu \leq 1420 {\rm MHz}$.

The Lyman-$\alpha$ system comprises of the diffuse  HI in the  dominantly ionized post-reionization inter galactic medium (IGM), which produces
distinct absorption features in the spectra of background QSOs. 
These absorption features known as the Lyman-$\alpha$ forest, provides one dimensional maps of the underlying HI fluctuation field along QSO
sight lines. The Lyman-$\alpha$ forest is known to be  a  powerful cosmological probe \citep{Croft_1999,Gnedin_2002, Mandelbaum_2003, Gallerani_2006, McDonald_2007,  Lesgourgues_2007, 2015AAS...22512503F}. The Baryon Oscillation Spectroscopic Survey (BOSS) \citep{Delubac_2015, FontRibera_2012} has measured the BAO imprint on the Lyman-$\alpha$ forest. 
The eBOSS \footnote{https://www.sdss.org/surveys/eboss/}  survey is expected to cover $6000$ square degrees and acquire data for $500,000$ quasars in the redshift range $0.8<z <3.5$.
The high number density of QSOs in this survey and the  high signal to noise ratio (SNR) measurement of the Lyman-$\alpha$ spectra allows 3D analysis \cite{McQuinn_2011, Sarkar_2015} and powerful investigation of the cosmological dark sector   \citep{Garzilli_2019, Viel_2003}.

Large scale numerical simulations indicate that on  large cosmological scales both the Lyman-$\alpha$ forest and the 
21-cm signal are biased tracers of the underlying dark matter (DM)
distribution \cite{Bagla_2010, Guha_Sarkar_2012, Sarkar_2016, Carucci_2017, Villaescusa_Navarro_2018}. This has allowed the possibility of studying cross-correlations between Lyman-$\alpha$ and post reionization 21-cm signal \cite{Guha_Sarkar_2010, Sarkar_2015, Carucci_2017, Sarkar_2018, Sarkar_2019}.
We consider the cross correlation of these post-reionization tracers with the weak-lensing convergence field to constrain dark energy models.
The cross-correlation of 21-cm signal and the Lyman-$\alpha$ forest with weak-lensing has been studied earlier \citep{GSarkar_2010, Vallinotto_2009}. We perform an improved analysis of the angular cross power spectrum for 
correlations of the  CMBR weak lensing convergence field with the Lyman-$\alpha$ forest and 21-cm signal  for constraining different dark energy parametrizations.
We use a visibility based realistic formalism for this purpose. The entire redshift range probed by the Lyman-$\alpha$ and the 21-cm signal is exploited in the present analysis whereby the signal is averaged over redshift bins. Cross-correlating the integrated Lyman-$\alpha$ absorption with weak-lensing convergence has been studied \citep{Vallinotto_2009}. We extend this idea to  the post-reionization signal whereby we consider the  signal stacked up over the redshift slices in the observed bandwidth. This is expected to yield greater SNR in detection and more stringent constraints on the dark energy parameters.

The paper is divided as follows. We first consider the dark energy models. We look at the auto correlation convergence power spectrum 
for CMBR weaklensing. We finally make signal to noise predictions with a Fisher matrix parameter estimation using a visibility based formulation of the cross-correlation of the Lyman-$\alpha$ forest flux and 21-cm signal with the convergence field. 

\section{Dark energy models}
The evolution of the Hubble parameter $H(a)$ for a spatially flat FRW Universe is given by 
\be
\frac{H(a)}{H_0}   = \sqrt{\Omega_{m_0}a^{-3}+ ( 1 - \Omega_{m_0} )  ~{\rm exp}\left[ -3\int_1^a da' \frac{1+w(a')}{a'}\right] }
\label{eq:hubble}
\ee
where $H_0$ and  $\Omega_{m_0}$ denote the  Hubble parameter and the  matter density parameter respectively at the present epoch and the Universe is assumed to be comprised of non-relativistic matter and a dark energy component with an evolving EoS $w(a)$.
We have used the cosmological parameters Planck18 results  \[ (\Omega_{m{_0}},\Omega_{b_{0}},H_0,n_s,\sigma_8, \Omega_K) = (0.315,~0.0496,~67.4,~0.965,~0.811, ~0)\]
from \citep{Planck2018} in this paper.

Our ignorance about the dynamics of Dark energy is modeled using the EoS parametrization $w(z)$ with $a = 1/(1+z) $.
There are innumerable possible choices for $w(z)$. However it has been shown that at most a two-parameter model can be optimally constrained from observations \cite{PhysRevD.72.043509}.
 
The model proposed by Chevallier Polarski \cite{CHEVALLIER_2001}  and   Linder \cite{PhysRevLett.90.091301} gave a phenomenological model-free parametrization to incorporate several features of dark energy. This model has been extensively used by the Dark Energy Task force 
\cite{2006astro.ph..9591A} as the standard  two parameter description of dark energy dynamics.  
The EoS is given by $w_{_{CPL}} (z) = w_0^{CPL} + w_a^{CPL}  \frac{z}{1+z}  $.
This model gives a smooth variation of $w(z) = w_0 + w_a$ at $ z \rightarrow \infty$ to $w(z) = w_0$ at $z = 0$.
It has also been shown that a wide class of quintessence scalar field models can be mapped into the CPL parametrization \cite{2015PhRvD..92d3001S} . 
However a better fit to both tracking and thawing class of models require a generalization of the CPL parametrization \cite{2016PhRvD..93j3503P}.
We use the following parametrizations in this work.
\begin{eqnarray}
w_{_{CPL}} (z) &= w_0^{CPL} + w_a^{CPL} \left ( \frac{z}{1+z} \right ) ~~~~~(CPL) \\
w_{_{7CPL}} (z) &= w_0^{7CPL}  + w_a^{7CPL} {\left ( \frac{z}{1+z} \right )}^{7} ~~~~~(7CPL)\\
w_{_{BA}} (z) &= w_0^{BA} + w_a^{BA} \left(\frac{z(1+z)}{1+z^2} \right)~~~~~(BA) 
\end{eqnarray}
Each model is characterized by two constant parameters $(w_0, w_a)$ with $w_a$ quantifying the evolution of dark energy from its present value set by 
$w_0$. The effect of dark energy EoS on observable quantities pertaining to the background cosmological evolution  and structure formation is discussed in the Appendix. 

\section{Weak-lensing convergence  power spectrum }

We consider the weak-lensing of the Cosmic microwave background radiation (CMBR). 
Gravitational lensing  deflects the photons which are free streaming from the last scattering
surface (epoch of recombination $z \sim 1000$ ) and manifests as a secondary anisotropy
in the CMBR  temperature maps. 
The effect of gravitational lensing can
be extracted from these maps by constructing various estimators for the convergence field $\kappa$  through quadratic combination of the $(T, E, B)$  fields \cite{Seljak_1999, Hu_2001}. Defining convergence $\kappa$ as $ \kappa = -\frac{1}{2} \nabla \cdot \mathbf{ \alpha}$ where, $\alpha$ is the total deflection, the convergence power spectrum for the CMBR weak lensing is given by 

\begin{equation}
C_{\kappa_{_{CMB}}}^{\ell}  = \frac{9}{4} \left(\frac{H_0}{c}\right)^4 \Omega^2_m{_{_0}} \int_0^{\chi_{rec}} \frac{ {g(\chi)}^2}{a^2{(\chi)}} P\left(\frac{\ell}{\chi},\chi\right) d\chi.
\end{equation}
If  $\chi_{rec} = \chi (z_{rec})$ is the comoving distance to the last scattering surface then  the weaklensing geometric kernel $g(\chi)$ is given by 
\be
g(\chi)  = \left( \frac { \chi_{rec} - \chi}{\chi_{rec}} \right )
\ee
We have incorporated the Limber approximation in the above expression.

\begin{figure}[h]
\begin{center}
\includegraphics[height=5cm]{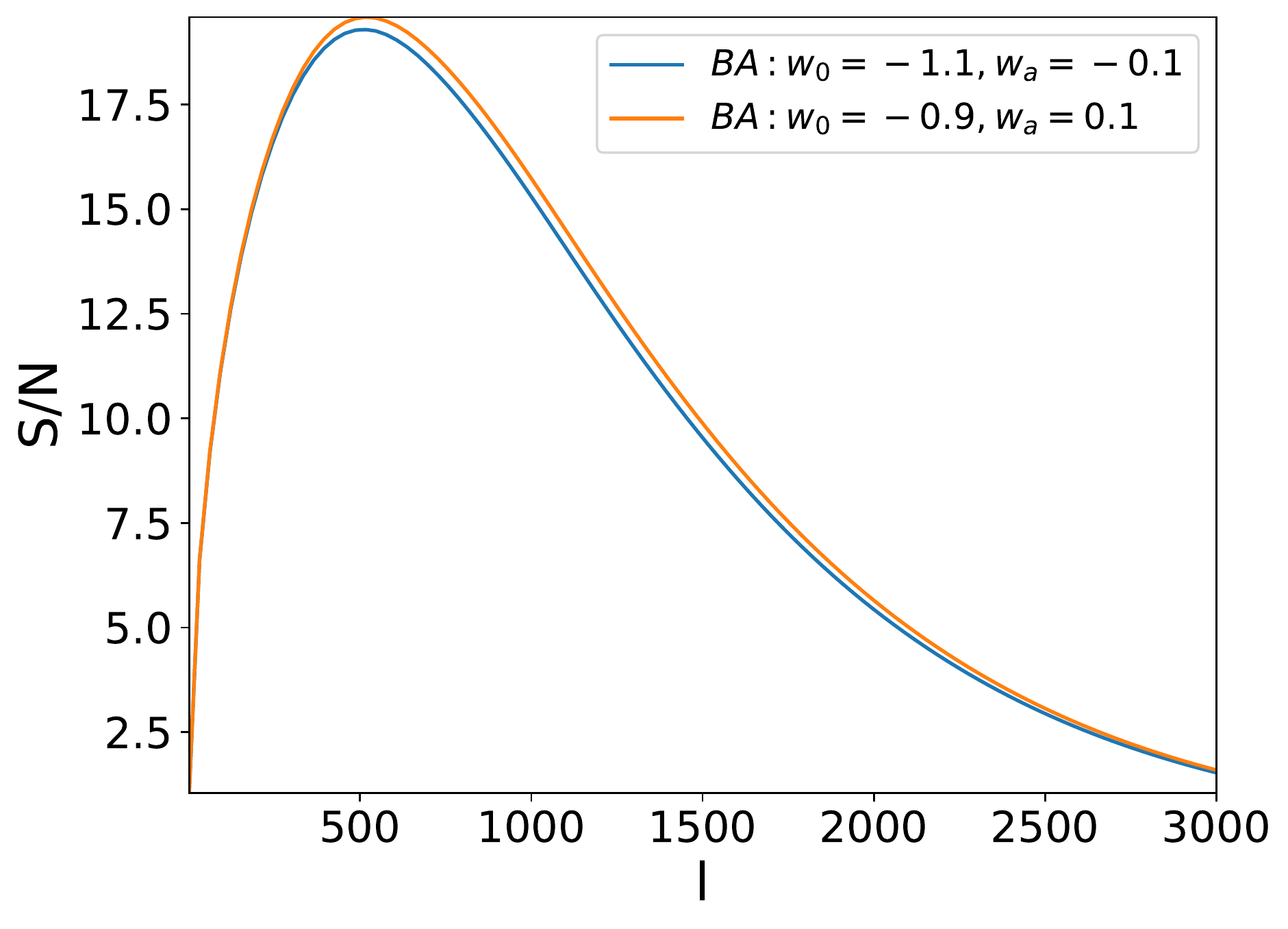}
\includegraphics[height=5cm]{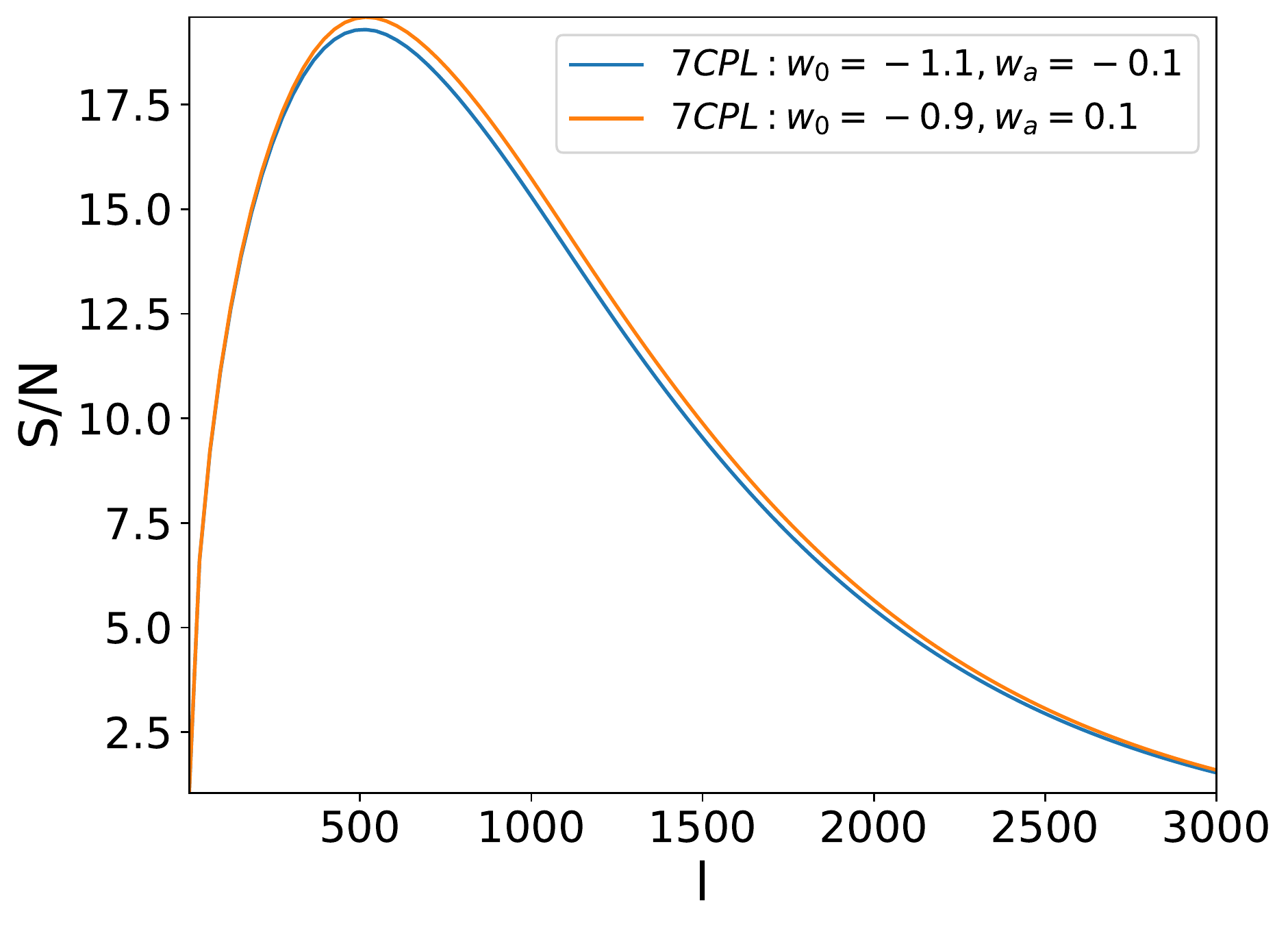}
\includegraphics[height=5cm]{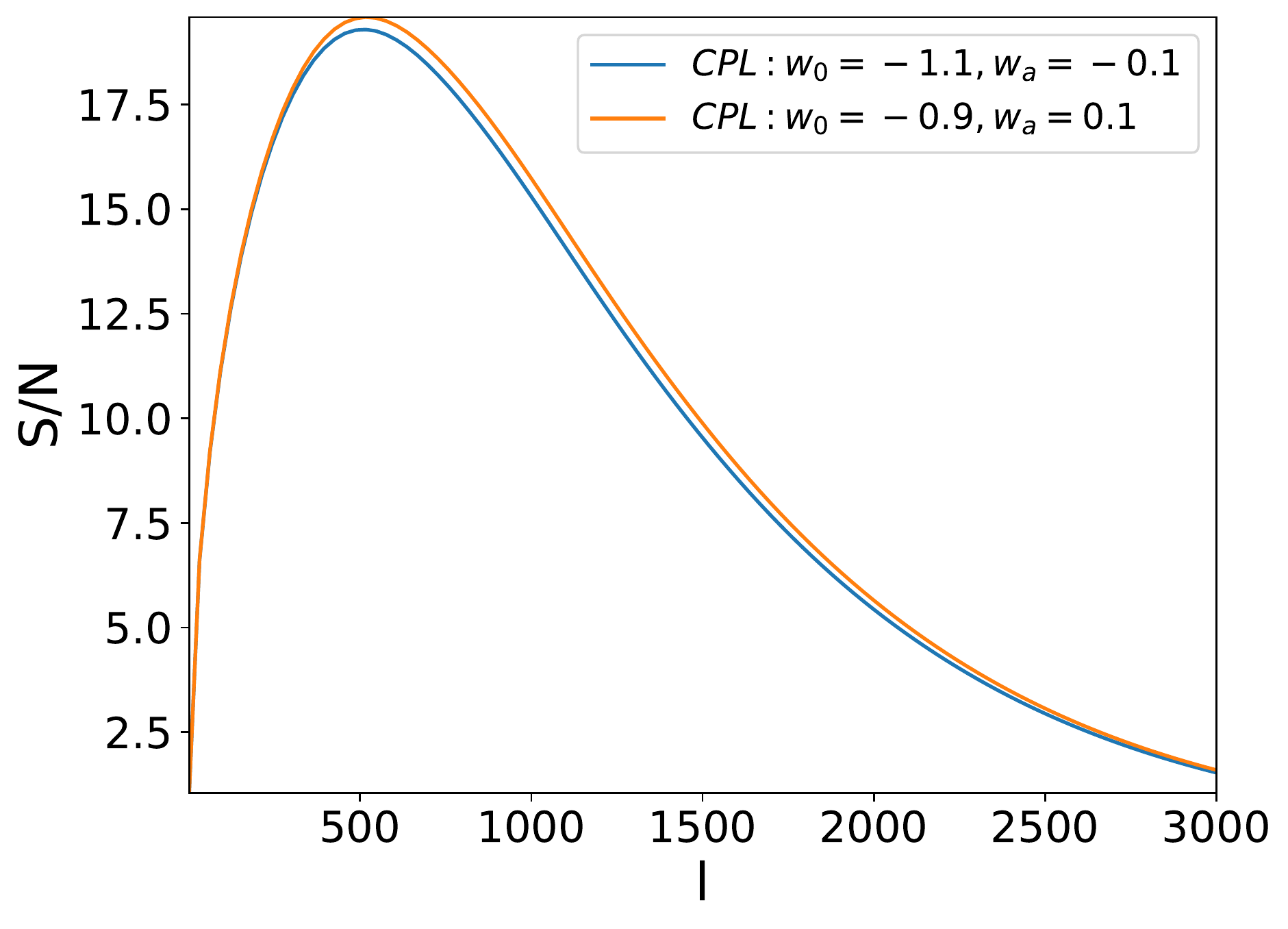}
\includegraphics[height=5cm]{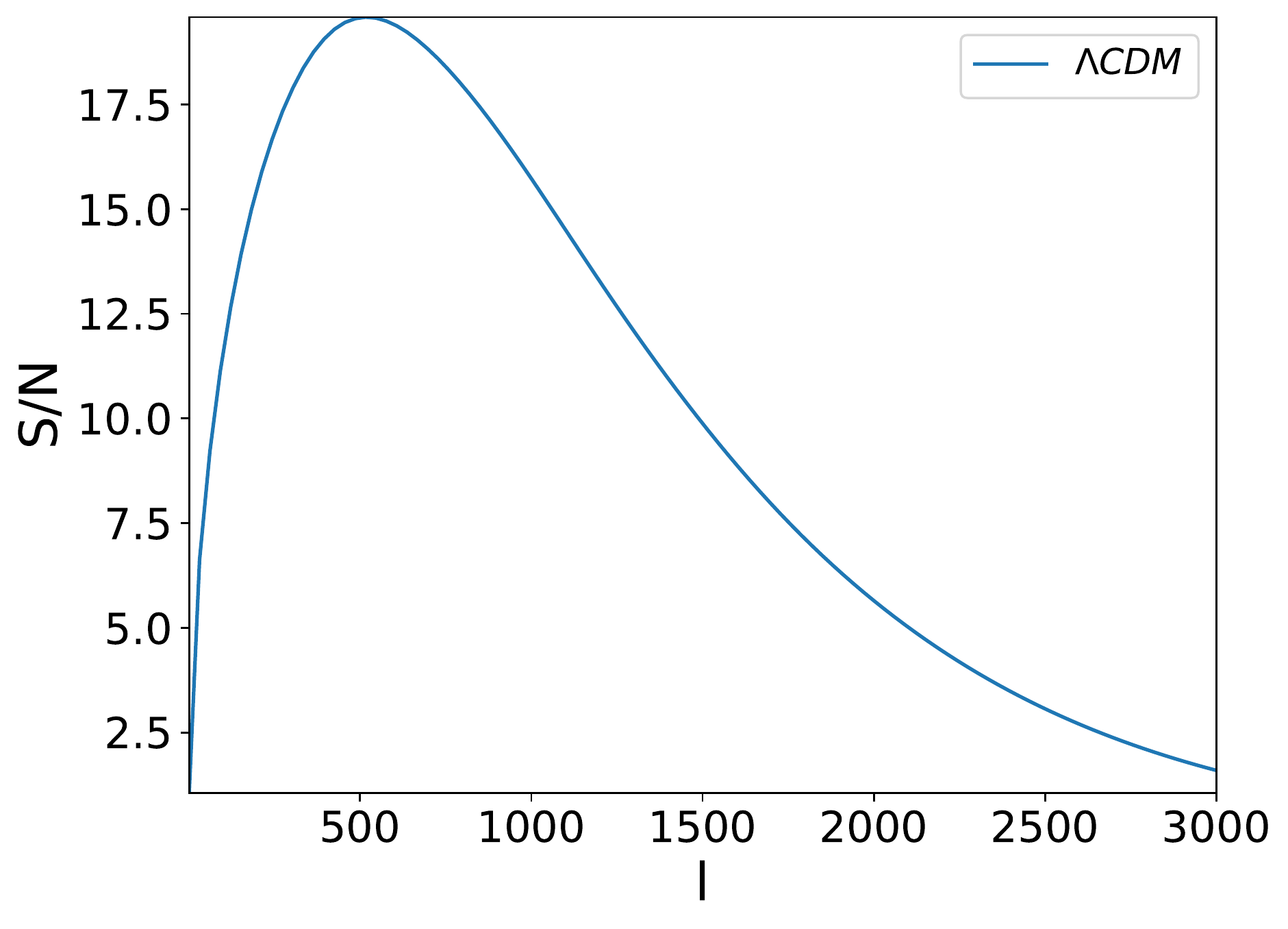}
\caption{ SNR for the  CMBR convergence power spectrum for different fiducial dark energy parametrizations. }
\label{fig:snrcmbr}
\end{center}
\end{figure}

 The Fisher matrix is diagonal for a full sky survey and the  noise for the CMBR convergence power spectrum is given by 
\begin{equation}
\Delta C_{\kappa_{_{CMB}}} ^{\ell} = \sqrt{\frac{2}{(2\ell+1)}} \left( C_{\kappa_{_{CMB}}}^{\ell}  + w^{-1} e^{{\ell}^2\sigma_b^2} \right)
\end{equation}
where the first term comes from cosmic variance and  the instrumental noise is encapsulated in the weight $w$ and a smoothing determined by the beam width $\sigma_b$. We may write  $w = (\sigma^2_{pix} \Omega_{pix})^{-1}$. Here 
$\sigma^2_{pix}$ is the error in each pixel which depends on the sensitivity $s$ and observation time for each pixel $t_{pix}$ as $\sigma_{pix} = s/\sqrt t_{pix}$. If the FWHM ( full width at half maximum) is denoted by $\theta_{fwhm}$ then $\Omega_{pix} = K \theta_{fwhm} \times \theta_{fwhm}$ and 
$\sigma^2_b = \theta^2_{fwhm}/8{\rm ln} 2$. The conversion factor $K^{-1} = \frac{1}{(2 \pi)^2} \int d^2 \vec \ell ~ \ell^2  C_{\ell}$, where $C_{\ell}$ denotes the CMB angular power spectrum converts noise in CMBR temperature angular power spectrum to that of convergence angular power spectrum. 
The factor $(2\ell +1)$ in the denominator counts the number of samples of  $C_{\kappa_{_{CMB}}} ^{\ell} $ for a given $\ell$. 
Figure (\ref{fig:snrcmbr}) shows the SNR for the CMBR weak-lensing angular power spectrum. We have assumed a CMBPol like experiment with pixel noise of  $\sigma^2_{pix} \approx 1\mu K$ and $\theta_{fwhm} = 3 {\rm arcmin}$ and $w^{-1} = 7.5   \mu K^{-2} {deg}^{-2} $  for our analysis \cite{Smith_2009}.

Figure (\ref{fig:snrcmbr}) shows the SNR for CMBR weak-lensing convergence angular power spectrum. The sensitivities for different dark energy EoS parameterizations differ by a few percent from the LCDM 
predictions. However, the sensitivities SNR $\sim 20$ is good enough to rule out some of the parametrizations at $3 \sigma$. But most EoS models remains degenerate to LCDM at these sensitivity levels since the difference of the angular power spectrum is $\sim 1 \%$.

\section{Cross correlation signal}

On large scales the Lyman-$\alpha$ forest and the redshifted 21-cm signal from the post reionization epoch are both known to be biased tracers of the underlying dark matter distribution.
We denote $\delta \F$ to denote the fluctuations in the Lyman-$\alpha$ transmitted flux $\delta \F = ( \F - \bar{\F} )/ \bar{\F}$. 
In the post reionization era the neutral fraction remains constant \citep{Lanzetta_1995, P_roux_2003, Noterdaeme_2009, Zafar_2013} and photo-ionization equilibrium leads to a power-law, temperature density relationship \citep{Hui_1997}.
Numerical simulations validate the  Fluctuating Gunn-Peterson effect \citep{Bi_1997, Viel_2002, Slosar_2009, Slosar_2011} whereby it is  reasonable to assume that on large scales the smoothed flux fluctuations  $\delta_{\F} \propto \delta$, where $ \delta$ denotes the dark matter overdensity field.

The HI 21-cm emission signal also arises  from the same redshift range as the Lyman-$\alpha$ forest.
However, they are sourced by the  DLAs which are believed to contain most of the HI during
the post reionization era.  The Lyman-$\alpha$ forest, on the contrary arises from the  low density HI
 in a predominantly ionized IGM.
 On large scales the HI 21-cm signal also traces the underlying dark matter  distribution. 
We use $\delta_{T}$ to denote the redshifted 21-cm brightness
temperature fluctuations.  

Including a redshift space distortion we may write both  $\delta_{\F}$ and $\delta_{T}$ in Fourier space as
\be
 \delta_{i}({\bf r}) = \int \ \frac{d^3
   {\bf{k}}}{(2\pi)^3} \ e^{i {\bf{k}}. {\bf r}}  \Delta_{i}({\bf{k}}) \,.
\label{eq:deltau}
\ee where $i={\F}$ and $T$ refer to the Ly-$\alpha$ forest
transmitted flux and 21-cm brightness temperature respectively, with  
\be 
\Delta_{i}({\bf{k}}) = \A_{i} [1 + \beta_{i} \mu^2]
\Delta({\bf{k}}) \, 
\ee 
where $\Delta({\bf{k}})$ is the dark matter
density contrast in Fourier space and  $\mu$ is the cosine of the angle
between the line of sight direction ${\bf \hat {n}}$ and the wave
vector ($ \mu = {\bf \hat{ k} \cdot \hat{n}}$). $\beta_{i}$ is the linear redshift 
distortion parameter. For the 21-cm signal  we have
\be 
\A_{T} = 4.0 \, {\rm {mK}} \,
b_{T} \, {\bar{x}_{\rm HI}}(1 + z)^2\left ( \frac{\Omega_{b0}
  h^2}{0.02} \right ) \left ( \frac{0.7}{h} \right) \left (
\frac{H_0}{H(z)} \right) 
\ee
where ${\bar{x}_{\rm HI}}$ is the mean neutral fraction. In the post-reionization epoch $z<6$,  $\Omega_{gas} \sim 10^{-3}$  and the  neutral hydrogen
fraction remains  with a value $ {\bar{x}_{\rm
    HI}} = 2.45 \times 10^{-2}$  \cite{Lanzetta_1995, P_roux_2003, Zafar_2013, Noterdaeme_2009}.  
We may write $\beta_T = f(z)/ b_T$ 
where $b_T$ denotes a bias and $f(z)$ is growth rate of density perturbations.  The bias function $b_T(k, z)$ is
 scale dependent below the Jeans scale. There is also additional scale dependence  
arising from the fluctuations in the ionizing background
\cite{poreion0}. The bias is also a monotonically growing function of redshift \cite{Guha_Sarkar_2012}. 
Several studies indicate that on large scales a constant linear bias model is reasonably valid \citep{Bagla_2010, Guha_Sarkar_2012, Sarkar_2016}.
The function $\beta_T(z)$ crucially imprints the dark energy parametrizations through its dependence of $f(z)$ and has a redshift dependence arising from both $f(z)$ and $b_T(z)$.

The interpretation of the linear distortion parameter, $\beta_{\F}$ for the Lyman-$\alpha$ forest, is different owing to the non-linear relation between the Lyman-$\alpha$ transmitted flux
and the underlying dark matter density field \cite{Slosar_2011}. Contrary to the parameters for the  HI 21-cm signal, the parameters
  $(\A_{\F}, \beta_{\F})$ are independent of each other and are sensitive to parameters like the IGM temperature-density relationship ($\gamma$) and the flux probability distribution function (PDF) of  the Lyman-$\alpha$ forest. 
  Analytical work \cite{2012JCAP...03..004S}  and extensive numerical simulations \cite{McDonald_2003, 2016JCAP...03..016C, Carucci_2017, 2020arXiv200909119F, 2019JCAP...02..050B} demonstrates that in the absence  of
  primordial non-gaussianity the  Lyman-$\alpha$ forest can be described by a linear theory with a scale independent bias on large scales. 
 
 We adopt approximate fiducial values $(\A_{\F}, \beta_{\F}) \approx ( -0.15, 1.11 )$ from the numerical simulations of Lyman-$\alpha$
 forest \cite{McDonald_2003}. The  bias function $b_T$ is taken from simulation results \cite{Guha_Sarkar_2012}. These  fiducial
values of the parameters are used for the Fisher matrix analysis.
 
  We consider the cross-correlation of post-reionization tracers with the weak lensing convergence field. 
 There is a crucial difference between the weak lensing field and the tracer fields. The former measures the integrated effect along the line of sight
 of a geometric kernel from redshift $z=0$ to the last scattering surface and, is hence sensitive to the fluctuations of the density field on  large scales. The tracers, namely the Lyman-$\alpha$ forest and redshifted 21-cm intesity maps on the contrary are tomographic probes of small scale fluctuations.  The cross-correlation thereby quantifies the evolution of small wavelength modes of the fluctuations on top of the long wavelength modes.
 Further, the noise and systematics which affect the auto-correlation signal appears only in the variance of the cross-correlation and may pose less challenge towards detection of the cross-correlation signal . The 21-cm signal is burried deep under galactic and extra-galactic foregrounds. Even after significant foreground removal, the cosmological origin of the 21 cm signal can only be ascertained only through a cross-correlation.

 \subsection{General formulation}
To formulate the cross-correlation angular power spectrum, we expand the convergence field in terms of spherical harmonics as
\be
\kappa({\bf \hat{n}}) = \sum_{\ell, m}^{\infty} a^{\kappa}_{\ell m} Y_{lm} ({\bf \hat{n}}) \label{eq:spherical} \ee
 The expansion coefficients $a^{\kappa}_{lm}$ can be obtained by inverting Eq.(\ref{eq:spherical}) as 
\be
a_{\ell m}^{\kappa} = \int d\Omega_{\n}~ \kappa(\n) Y_{\ell m}^*({\n}) 
\ee
Thus, on using the expression for $\kappa(\n)$ we have 
\be
a_{\ell m}^{\kappa} = \int d\Omega_{\n}~  Y_{\ell m}^*({\n}) \frac{3}{2} \left(\frac{H_0}{c}\right)^2 \Omega_m{_{_0}} \int_0^{\chi_s}  g(\chi) ~\chi~ \frac{\delta(\chi \n, \chi )}{a(\chi)} d\chi
\ee
Writing 
\be
 \delta( \chi \n, \chi) = \int  \frac{d^3 {\bf{k}}}{(2\pi)^3} \ e^{i {\k}.{\n} \chi} \Delta({\bf{k}}) D_{+} (\chi)
\ee 
and using the Raleigh expansion
\be
 e^{i{\mathbf{k}}\cdot {\mathbf{n}} \chi }  = 4\pi \sum_{\ell,m}{
   (-i)}^{\ell} j_{\ell} (k \chi )Y_{\ell m}^*({\bf{\hat{k}}}) Y_{\ell m}({\bf{\hat{n}}})
\ee
along with the normalization
\be 
\int d\Omega_{\n}~ Y_{\ell m}^*({\n}) Y_{\ell m}({\n}) = 1  \ee
we have
\be
a_{\ell m}^{\kappa} =  4 \pi    {(-i)}^{\ell} \int  \frac{d^3 {\bf{k}}}{(2\pi)^3}    \int_0^{\chi_s}  d \chi~ \A_{\kappa} (\chi) D_{+} (\chi)  j_{\ell} (k \chi )  \Delta({\bf{k}}) Y_{\ell m}^*({\bf{\hat{k}}})
 \ee
where
\be \A_{\kappa} (\chi) = \frac{3}{2 } \left(\frac{H_0}{c}\right)^2 \Omega_m{_{_0}}\frac{    g(\chi) ~\chi~} {a(\chi)} \ee
For the post reionization tracers Lyman-$\alpha$ and redshifted 21-cm signals we define two fields on the sky by integrating $\delta_i(\chi \n , \chi)$ along the radial direction 
\be 
F_i(\n) = \frac{1}{\chi_2 - \chi_1}\sum_{\chi_1}^{\chi_2}  \Delta \chi ~ \delta_i(\chi \n , \chi) 
\label{eq:average}
 \ee
Previous works \cite{GSarkar_2010, tanaka2020detectability} consider these fields at a given redshift, where the radial information is retained for tomographic study.
The weak-lensing convergence, on the contrary consists of a line of sight integral whereby the redshift information is lost.
We consider an average over the signals from redshift slices and thus lose the redshift information but improve the SNR when cross-correlating with the 
weak-lensing field,

The  expansion coefficients for Lyman-$\alpha$ and redshifted 21-cm signals can be generally written as 
\be 
a_{\ell m}^{F_i} = 4 \pi {(-i)}^{\ell} \int  \frac{d^3 {\bf{k}}}{(2\pi)^3} \frac{1}{\chi_2 - \chi_1}\sum_{\chi_1}^{\chi_2}  \Delta \chi ~   ~\A_i (\chi) D_{+} (\chi) \left ( j_{\ell} (k \chi ) - \beta_i \frac{d^2 j_{\ell}(k \chi) }{ d {(k \chi)}^2 } \right ) \Delta({\bf{k}}) Y_{\ell m}^*({\bf{\hat{k}}}) 
  \ee
Defining the cross-correlation angular power spectrum as
\be 
\langle a_{\ell m}^{\kappa} {a^{F_i *}_{\ell'  m' }}\rangle = 
C_{\ell}^{\kappa F_i}  \delta_{\ell, \ell'} \delta_{m, m'} \ee
we have
\be
C_{\ell}^{\kappa F_i}  =\frac{2}{\pi(\chi_2-\chi_1)}  \int_{0}^{\chi_s} d\chi  \sum_{\chi_1}^{\chi_2} \Delta \chi'   \A_{\kappa} (\chi) D_{+} (\chi) A_{i } (\chi') D_{+} (\chi')  \int dk ~k^2  j_{\ell}( k \chi) J_{\ell} (k \chi') P(k) 
\ee
where
\be
J_{\ell}( k \chi) =  \left ( j_{\ell} (k \chi ) - \beta_i \frac{d^2 j_{\ell}(k \chi) }{ d {(k \chi)}^2 } \right ) \ee
Similarly the auto-correlation angular power spectra may be written as
\be
C_{\ell}^{\kappa \kappa }  =\frac{2}{\pi}  \int_{0}^{\chi_s} d\chi  \int_{0}^{\chi_s} d \chi'   \A_{\kappa} (\chi) D_{+} (\chi) A_{\kappa} (\chi') D_{+} (\chi')  \int dk ~k^2  j_{\ell}( k \chi) j_{\ell} (k \chi') P(k) 
\ee
and
\be
C_{\ell}^{F_i F_i }  =\frac{2}{\pi( \chi_2 - \chi_1)^2}  \sum_{\chi_1}^{\chi_2} \Delta \chi  \sum_{\chi_1}^{\chi_2} \Delta \chi'   \A_{i} (\chi) D_{+} (\chi) A_{i} (\chi') D_{+} (\chi')  \int dk ~k^2  J_{\ell}( k \chi) J_{\ell} (k \chi') P(k) 
\ee

\subsection{Visibility based approach in ``flat-sky" approximation}
Radio interferometric observations of the redshifted 21-cm signal directly measures the complex Visibilities which are the Fourier components of the intensity distribution on the sky.
The radio telescope typically has a finite beam which allows us to use the "flat-sky" approximation.
Instead of expanding the fields $\kappa$ and $\delta_i$ in the basis of spherical harmonics, we shall now obtain a simplified expression for the angular power spectrum by considering the flat sky approximation whereby we can use the Fourier basis.
We define Visibilities as
\be
V_{F_i}( \vu )  = \int d^2 \vtheta ~  a(\vtheta) F_i ( \vtheta ) ~e^{ -2\pi i \vu . \vtheta} \ee
\be
V_{\kappa}( \vu )  = \int d^2 \vtheta ~  \kappa ( \vtheta ) ~e^{ -2\pi i \vu . \vtheta} \ee
where $a(\vtheta)$ denotes the beam function of the telescope measuring the  angular coverage of the 21-cm or Lyman-$\alpha$ forest survey.
Thus
\begin{eqnarray}
V_{F_i}( \vu )  = \frac{1}{\chi_2- \chi_1} \sum_{\chi_1}^{\chi_2}  \Delta \chi \int d^2 \vtheta ~  \A_i(\chi) \int \frac{d^2 \kperp d \kpar}{(2\pi)^3}  [ 1 + \beta_i(\chi)  \mu^2 ] \Delta({\bf k} ) D_{+}(\chi) \nonumber \\ \times  e^{ i \kpar \chi} ~ a(\vtheta)~ e^{  i ( \kperp \chi     -2\pi  \vu ). \vtheta} 
\end{eqnarray}
where $ \mu = \kpar/k$. Performing the $\vtheta$ integral, we have
\begin{eqnarray}
V_{F_i}( \vu )  = \frac{1}{\chi_2- \chi_1} \sum_{\chi_1}^{\chi_2}  \Delta \chi ~  \A_i(\chi) \int \frac{d^2 \kperp d \kpar}{(2\pi)^3}  [ 1 + \beta_i(\chi)  \mu^2 ] \Delta({\bf k} ) D_{+}(\chi)  \nonumber \\ \times  e^{ i \kpar \chi} ~  \widetilde{a}_{i}  \left ( \frac{\kperp \chi }{2 \pi}     - \vu \right )
\end{eqnarray}
Where the aperture function $\widetilde{a}_{i} (\vu)$ is the  Fourier transformation of the telescope beam function $ a(\vtheta)$.
Similarly, for the convergence field we have 
\begin{eqnarray}
V_{\kappa}( \vu )  =  \int_{0}^{\chi_s}  d \chi ~  \A_\kappa(\chi) \int \frac{d^2 \kperp d \kpar}{(2\pi)^3}  \Delta({\bf k} ) D_{+}(\chi)  ~  e^{ i \kpar \chi} ~   \delta_D  \left ( \frac{\kperp \chi}{2\pi}     -  \vu \right )
\end{eqnarray}
where, we have assumed an almost full sky weak-lensing survey.
We are interested in the Visibility-Visibility correlation 
\begin{eqnarray}
\langle V_{F_i}( \vu )  V^{*}_{\kappa}( \vu' ) \rangle = \frac{1}{\chi_2- \chi_1} \sum_{\chi_1}^{\chi_2}  \Delta \chi  \int_{0}^{\chi_s}  d \chi' ~  \A_i(\chi) \A_\kappa(\chi') D_{+}(\chi) D_{+}(\chi') 
\int \frac{d^2 \kperp d\kpar}{(2 \pi)^3}   e^{ i \kpar (\chi -  \chi')} \nonumber \\ \left [ 1 + \beta_i(\chi)  \frac{\kpar^2}{\kpar^2 + \kperp^2} \right ]  \widetilde{a}_i \left ( \frac{\kperp \chi}{2 \pi} - \vu \right ) \delta_D  \left ( \frac{\kperp \chi'}{2\pi}     -   \vu' \right )  P(k) \nonumber \\
\end{eqnarray}
Defining $ C^{F_i \kappa}(U) = \langle V_{F_i}( \vu )  V^{*}_{\kappa}( \vu' ) \rangle $
\begin{eqnarray}
C^{F_i \kappa}(U) = \frac{1 }{\pi(\chi_2- \chi_1)} \sum_{\chi_1}^{\chi_2}  \Delta \chi  \int_{0}^{\chi_s} \frac{ d \chi'}{\chi'^2}  ~  \A_i(\chi) \A_\kappa(\chi') D_{+}(\chi) D_{+}(\chi') 
\int_0^{\infty} d\kpar  \cos \kpar (\chi -  \chi')  \nonumber \\ \left [ 1 + \beta_i(\chi)  \frac{\kpar^2}{\kpar^2 + \left (\frac{2 \pi \vu}{\chi'} \right )^2} \right ]  \widetilde{a}_i \left ( \frac{ \chi - \chi'}{\chi'} \vu\right )  P \left (\sqrt{\kpar^2 + \left (\frac{2 \pi \vu}{\chi'} \right )^2 } \right ) \nonumber \\
\end{eqnarray}
If the aperture function $\widetilde a_i$ is peaked,  we may approximately write this as
\begin{eqnarray}
C^{F_i \kappa}(U) = \frac{1 }{\pi(\chi_2- \chi_1)} \sum_{\chi_1}^{\chi_2}  \frac{\Delta \chi}{\chi^2}    ~  \A_i(\chi) \A_\kappa(\chi) D_{+}(\chi)^2
\int_0^{\infty} d\kpar   \left [ 1 + \beta_i(\chi)  \frac{\kpar^2}{k^2} \right ]   P (k) \nonumber \\
\end{eqnarray}
with $k = \sqrt{\kpar^2 + \left (\frac{2 \pi \vu}{\chi'} \right )^2 } $.

The auto-correlation angular power spectrum may be similarly written as 
\begin{eqnarray}
C^{F_i F_i}  (U) = \frac{1 }{\pi(\chi_2- \chi_1)^2} \sum_{\chi_1}^{\chi_2}  \frac{\Delta \chi}{\chi^2}    ~  \A_i(\chi)^2  D_{+}(\chi)^2
\int_0^{\infty} d\kpar   \left [ 1 + \beta_i(\chi)  \frac{\kpar^2}{k^2} \right ] ^2  P (k) 
\end{eqnarray}
and
\begin{eqnarray}
C^{\kappa \kappa}  (U) = \frac{1 }{\pi} \int_{0}^{\chi_s}  \frac{d \chi}{\chi^2}    ~  \A_\kappa(\chi)^2  D_{+}(\chi)^2
\int_0^{\infty} d\kpar    P (k) 
\end{eqnarray}

We note here that working in the Fourier basis necessarily makes the signal nonergodic when one looks at correlation between two time slices (due to time evolution of all the relevant quantities). Further, one also notes the inseparability of the baseline $\bf U$ (transverse) from the frequency (radial) in this formalism \cite{Sarkar_2017}.

The angular power spectrum for two redshifts separated by $\Delta$ is known to decorrelate very fast in the radial direction \cite{poreion7}.
In this work we consider the summation in Eq (\ref{eq:average}) to extend over redshift slices whose separation is more than the typical decorrelation length. This 
ensures that in the computation of noise each term in the summation may be thought of as an independent random variable and the mutual covaraiances between the slices may be ignored.
This gives us the SNR in the measurement of $C^{F_i \kappa}(U)$ as
\be
 \frac{C^{\F_i \kappa} }{\sigma_{_{\F_i \kappa}}}= \frac{ C^{\F_i \kappa} \sqrt{2\ell + 1} \sqrt{N_c} } {\sqrt{(C^{\kappa \kappa} + \langle N^{\kappa } \rangle )( C^{F_i F_i} + \langle N^{F_i} \rangle )}}\ee
where $N_c$ is the number of redshift slices over which the average in Eq (\ref{eq:average}) is taken and $ \langle N^{F_i} \rangle $ and $\langle N^{\kappa}\rangle $ denotes the average of the noise power spectrum for $F_i$ and $ \kappa$ respectively.
This variance  is used in the Fisher matrix analysis  for constraining various dark energy EoS parameters.

The parameters  $\A_{\F} $, $\beta_{\F} $, $b_T$ and $\bar x_{HI}$ along with the cosmological parameters model the cross-correlation signal.
However the parameters $\A_{\F} $, $\beta_{\F} $, $b_T$ and $\bar x_{HI}$ are largely uncertain.
We perform the Fisher matrix analysis  assuming $\A_\F$, $\beta_\F$, $\A_T$ (related to $\bar x_{HI}$)  and $\beta_T$ (related to $b_T$)  along with the DE EoS parameters 
$w_0$ and $w_a$ to be the free parameters. 
The Fisher matrix is  given by 
\be
{\rm \mathbf F} _{a b} = \sum_\ell\frac{1}{\sigma_{_{F_i \kappa}}^2 }\frac{\partial C_{\ell} ^{F_i \kappa} }{\partial q_a } \frac{ \partial C_{\ell}^{F_i \kappa}}{\partial q_b}
\label{eq:Fisher}
\ee
where we have $q_a = ( w_0, w_a, \A_\F, \beta_\F, \A_T, \beta_T )$. The parameters $( w_0, w_a)$ are different for different models. 
The Cramer Rao  bound gives the errors on the $a^{th}$ parameter $ \delta q_a  = \sqrt{  {\rm \mathbf F}^{-1}_{aa}}$. 
The error projections on $ (w_0, w_a)$ are obtained for different dark energy models by marginalizing over the other parameters.

\section{Cross-correlation of CMBR weak-lensing with Lyman-$\alpha$ forest}
The redshift distribution of quasars is known to peak in  the redshift range $1.5\leq z \leq 3$ \cite{Schneider_2005}. 
In our work we consider quasars in these redshifts only and also consider that the Lyman-$\alpha$ forest spectrum for these quasars are measured at  a  high SNR.
We note that for any quasar the region $10, 000~ {\rm km~ sec^{-1}}$  blue-wards of the quasar's emission 
line is contaminated by the quasar's proximity
effect and the Stromgen sphere. We exclude this from the Lyman-$\alpha$ forest spectra.
We also note that pixels at least $1,000 ~{\rm{ km ~sec}^{-1}}$
red-ward of the quasar's needs to be excluded from the spectra to avoid contamination from  Lyman-$\beta$ forest and O-VI lines.
Thus, for a  quasar at the fiducial redshift $z_Q =  2.5$, the Lyman-$\alpha$ forest can be  measured in the redshift range $ 1.96 \leq z \leq 2.39$ 
covering a band $z_2 - z_1 = 0.43$.
The BOSS survey of $100,000$ square degrees, measures quasar spectra of at least $150,000$ quasars in the redshift range $2.15 < z < 3.5$ which can be used for cross-correlation \cite{2013AJ....145...10D}.
We note that for the Lyman-$\alpha$ forest at a redshift $z$ the noise
is given by 
\be 
N^{\cal F} = \frac{1}{\bar n_Q}  \left ( \sigma^2_{\F L}  +  \sigma^2_{\cal F N} \right ) \ee
where
\[\sigma^2_{\F L} =   \int d ^2 {\mathbf U} ~~C^{\F \F}  (U) \]

and $\bar n_Q$ is the angular number density of quasars. The quantity  $\sigma^2_{\cal F N}$ is a pixel noise contribution.
The observations of Lyman-$\alpha$ forest \cite{coppolani, D_Odorico_2006}
shows that for the Lyman-$\alpha$ forest smoothed over $\sim 50 {\rm Km s}^{-1}$, the variance of the flux fluctuations has a value $\sigma^2_{\F L} \approx 0.02$.
We adopt this value by choosing the corresponding smoothing scale. 
We also assume an average $S/N = 5$  for every pixel in the spectra used for cross-correlation. This gives us
$ \sigma^2_{\cal F N} = 0.04 \bar \F(z) ^{-2} \left ( \frac{4.6 \times 10^{-4} } {\Delta z } \right ) $. The mean flux $\bar \F(z) \sim 1$.  The variation of $\bar \F(z)$ is however noted \cite{Becker}
The averaging over $N_c = \frac{z_2 - z_2}{\Delta z }$ different slices of redshift gives $\langle N^{\F} \rangle$.
The smoothing scale $\Delta z$ is chosen by noting that at a  given redshift $z$ the signal decorrelates over  $\Delta z \sim 10^{-3} (1 + z)^2 \left ( \frac{\ell}{100} \right ) ^{-.0.7} $ \cite{poreion7}.
\begin{figure}[h]
\begin{center}
\includegraphics[height=6cm]{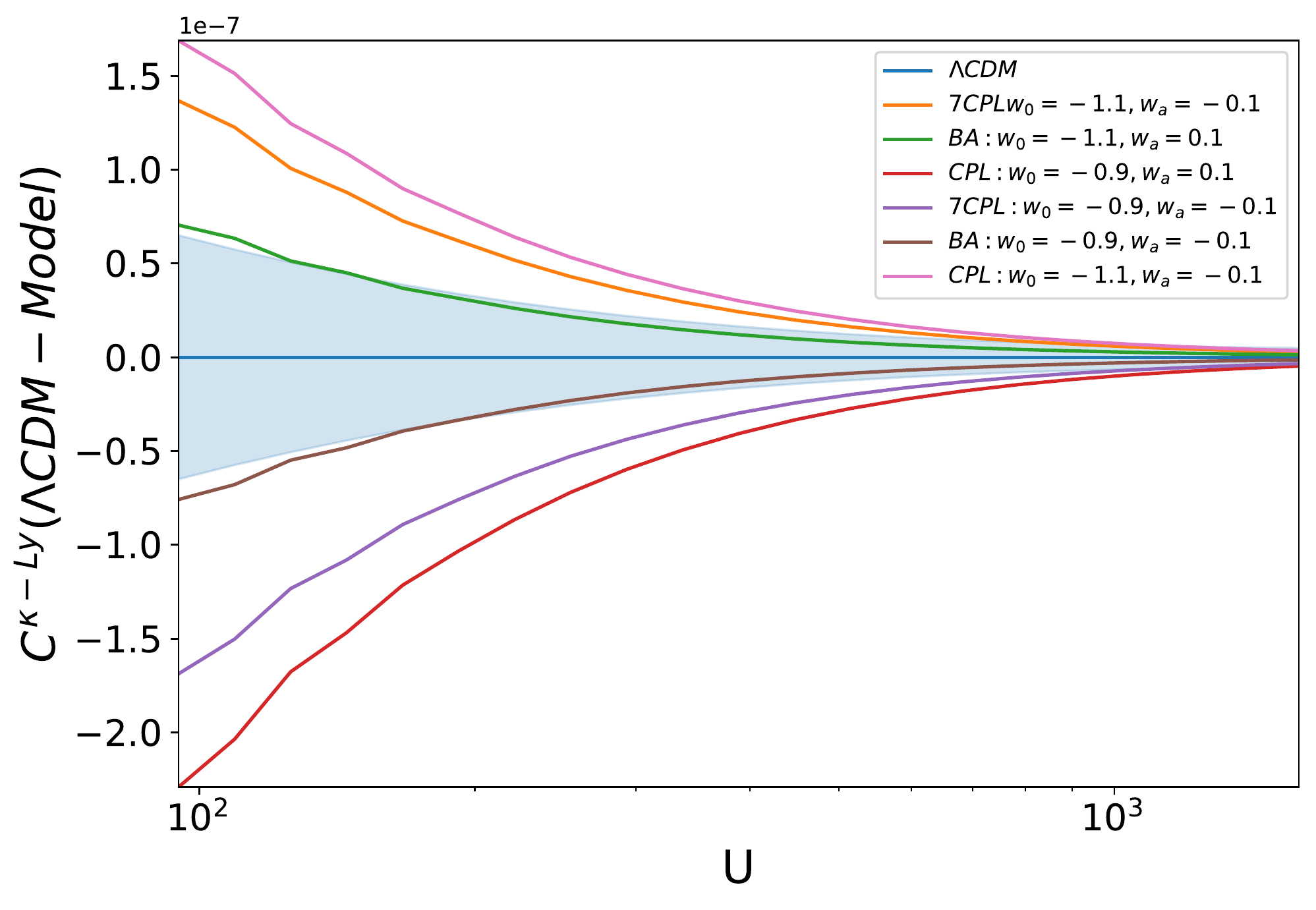}
\includegraphics[height=6cm]{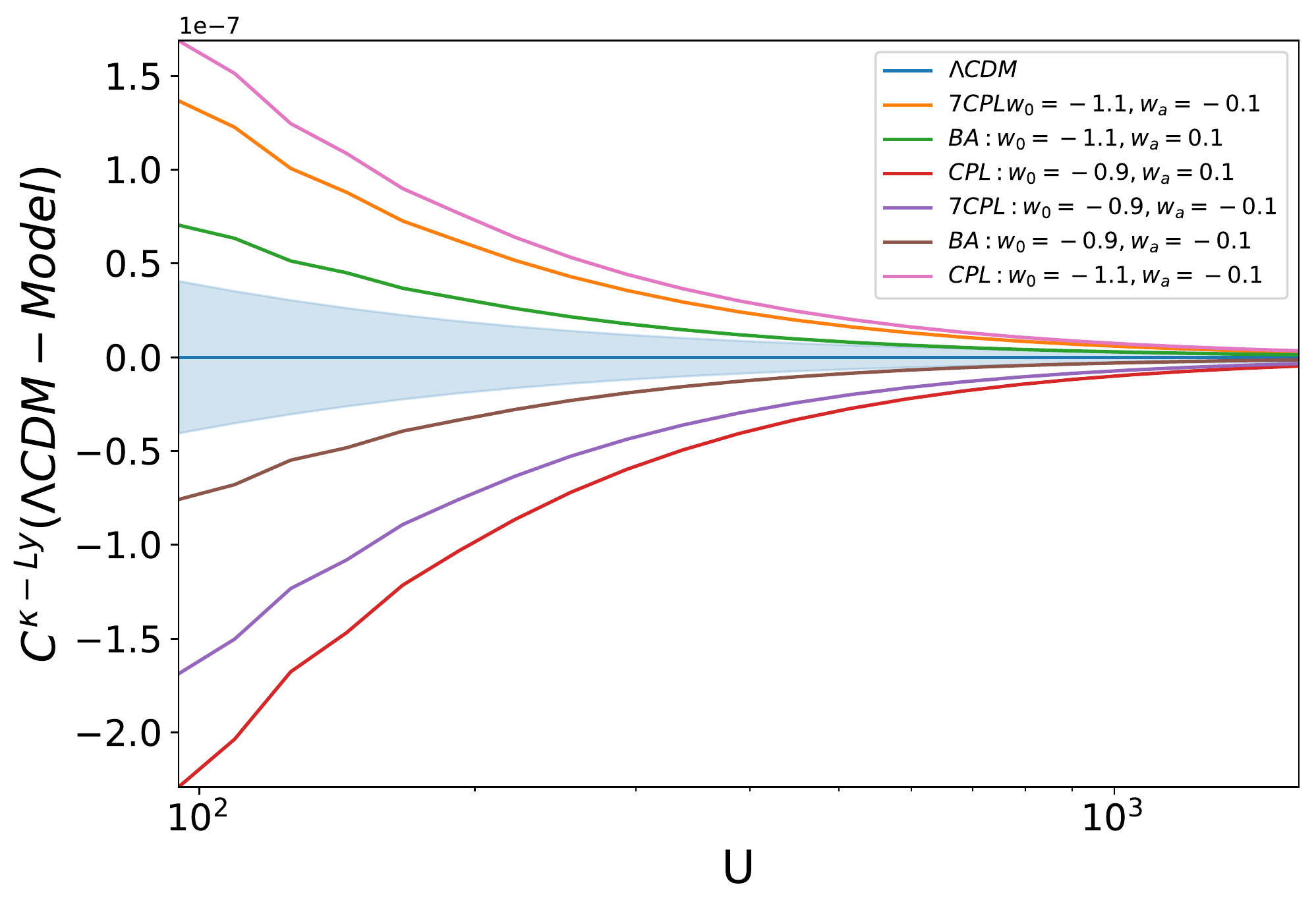}
\includegraphics[height=6cm]{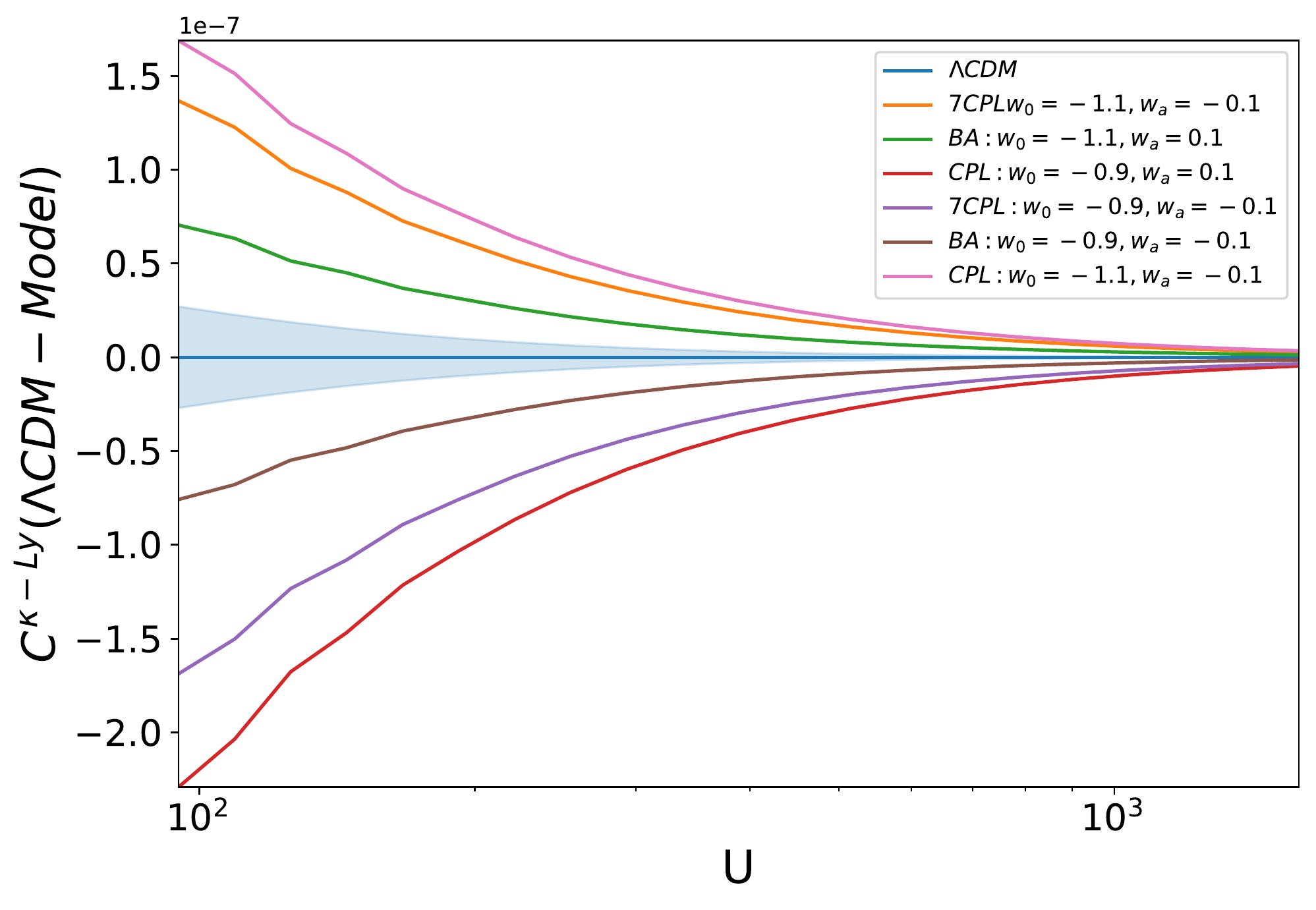}
\caption{ The  difference of the  Lyman-$\alpha$ forest - CMBR convergence cross power spectrum for different fiducial dark energy parametrizations
from the  the $\Lambda CDM$ model. The $1-\sigma$ error is shown by the shaded region. The first and the second figure corresponds to a quasar number density of $\bar{n}_Q = 16 {\rm deg}^{-2}$  and  $\bar{n}_Q = 64 {\rm deg}^{-2}$ respectively. The last figure shows the best case scenario in the cosmic variance limit of no observational noise. }
\label{fig:signal-cross-Ly}
\end{center}
\end{figure}
\begin{figure}[h]
\begin{center}
\includegraphics[height=6cm]{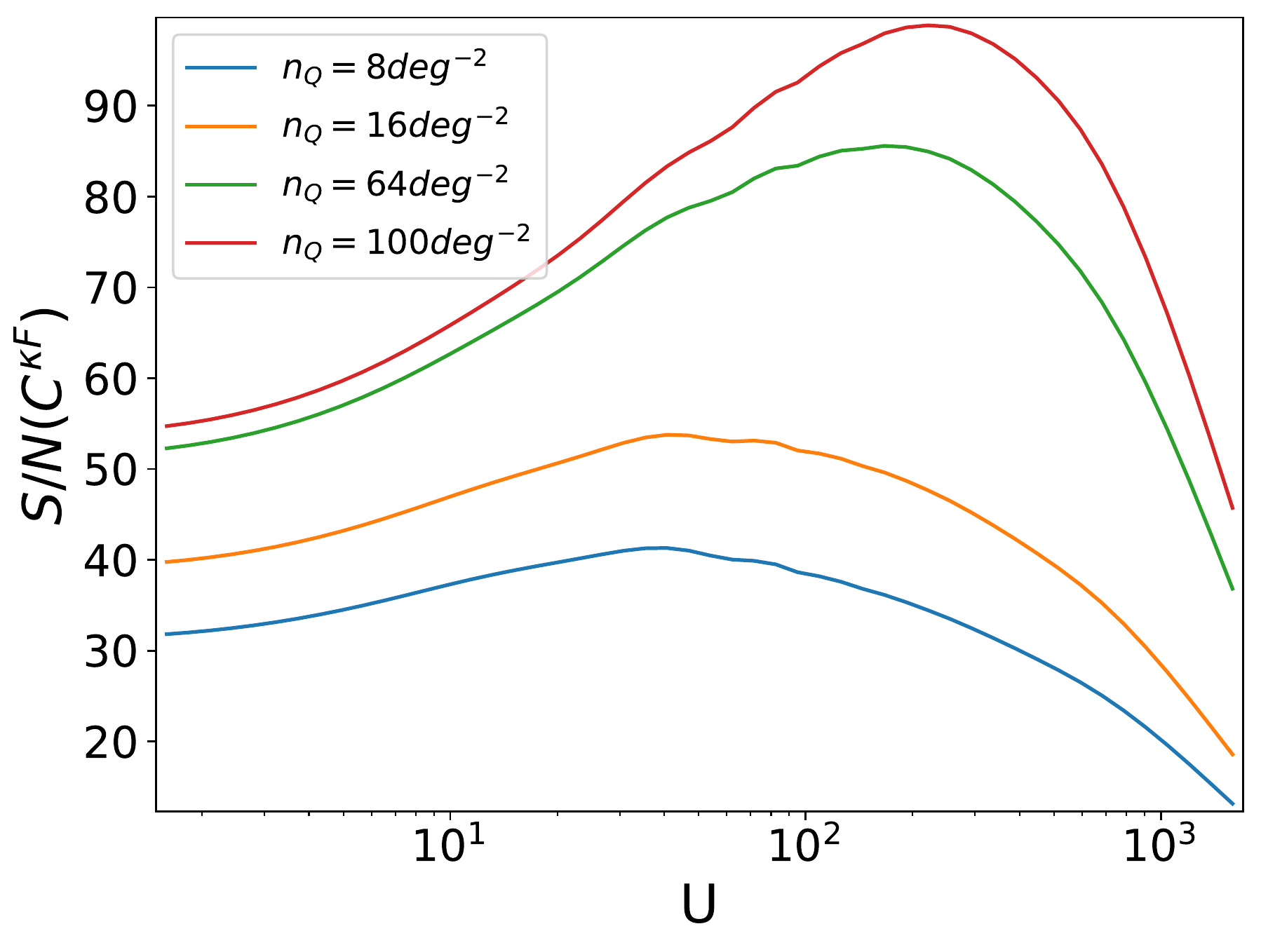}
\caption{ Signal to Noise ratio for the Convergence-Lyman-$\alpha$ angular power spectrum for the fiducial LCDM model for $\bar{n}_Q = 8, 16, 64, 100 ~{\rm deg}^{-2}$ respectively. }
\label{fig:SNRLy}
\end{center}
\end{figure}
Figure (\ref {fig:signal-cross-Ly}) shows the difference of the  Lyman-$\alpha$ CMBR convergence cross angular power spectrum 
for different dark energy EoS parametrizations from the cross angular power spectrum for the LCDM model.
 We consider Planck sensitivies for the cross-correlation with  $\sigma_{pix} \Omega_{pix}^{1/2}  = 0.7 \mu K.deg$. 
 For Lyman-$\alpha$ forest we have considered a BOSS like experiment with $\bar n_Q = 8, 16, ~  { \rm  and}~64 {\rm deg}^{-2}$ for our analysis.
 
We find that for $\bar n_Q = 16 {\rm deg}^{-2}$ with LCDM fiducial model a peak SNR of  $\sim 50$ (see Fig (\ref{fig:SNRLy})) is obtained at $ \ell \sim 300$. This allows CPL and 7CPL models to be differentiable from the LCDM model at $ > 3 \sigma $ sensitivity for $\ell > 1000$. The BA model however remains at $1\sigma$ from the LCDM and can not be differentiated at these sensitivities.
However its variance of the cross-correlation signal  is very sensitive to the quasar sampling and dense sampling is expected to improve the sensitivity  level for  the cross-correlation.
A Lyman-$\alpha$ survey with $\bar n_Q = 64 {\rm deg}^{-2}$ the cross correlation has a  peak SNR of $\sim 80$  at $ \ell \sim 1200$ [see Fig (\ref {fig:SNRLy})] . For such sensitivities 
CPL and 7CPL models can be differentiated at  $ > 5  \sigma$ for $ \ell > 1500$.
The best case scenario is shown in the last in figure which at $\bar n_Q \sim 100 {\rm deg}^{-2}$. 
\begin{figure}[h]
\begin{center}
\includegraphics[height=5.125cm]{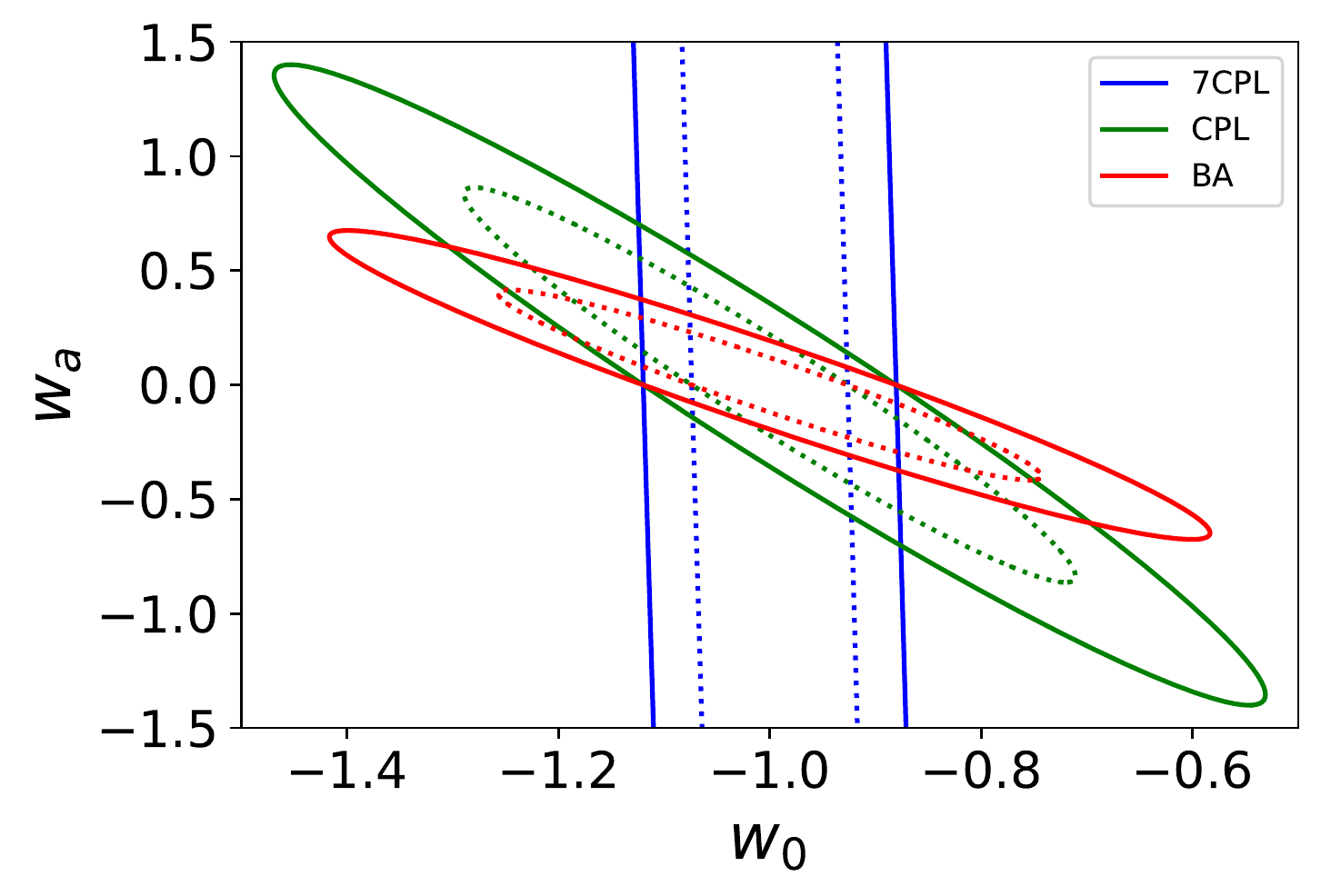}
\includegraphics[height=5.125cm]{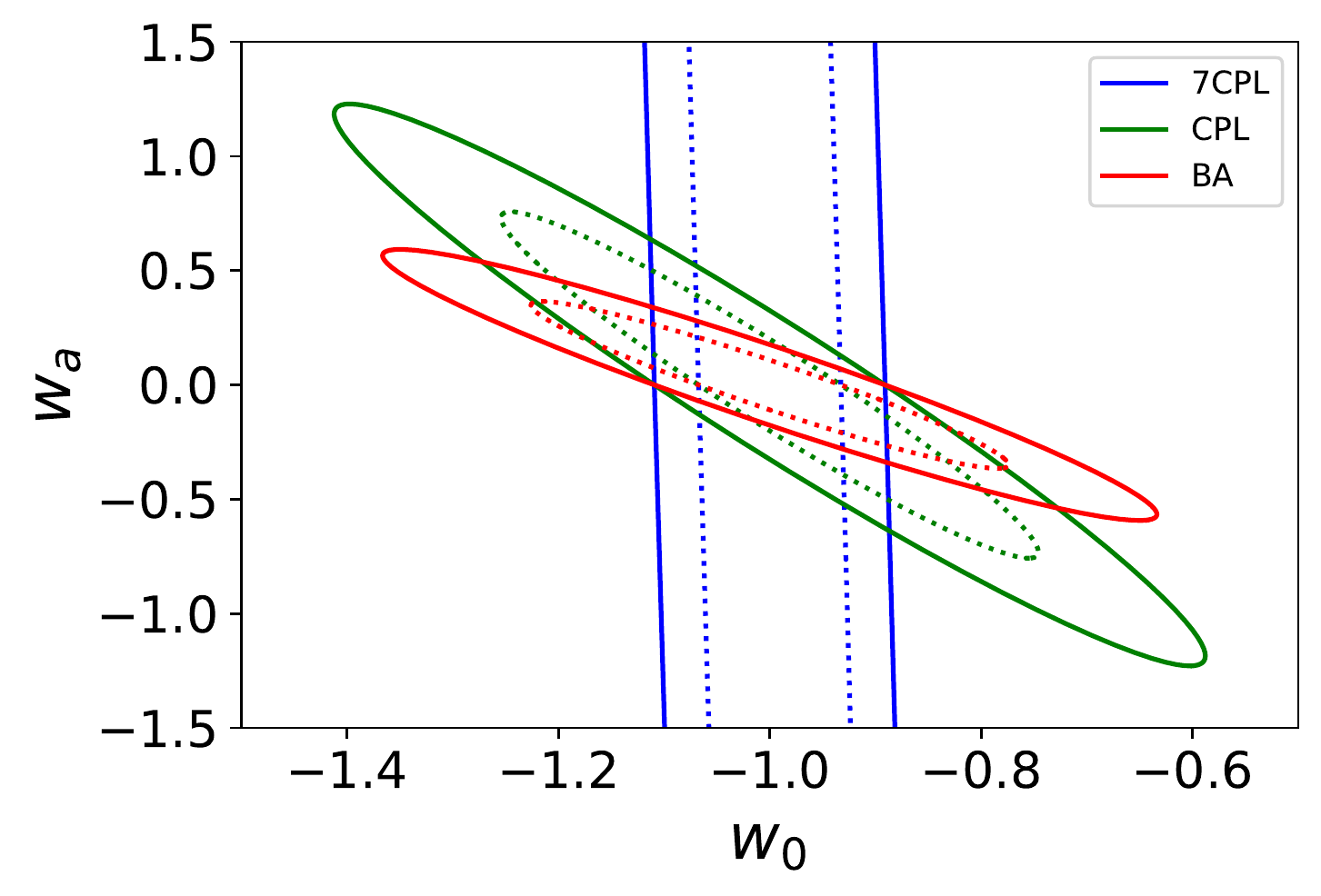}
\includegraphics[height=5.125cm]{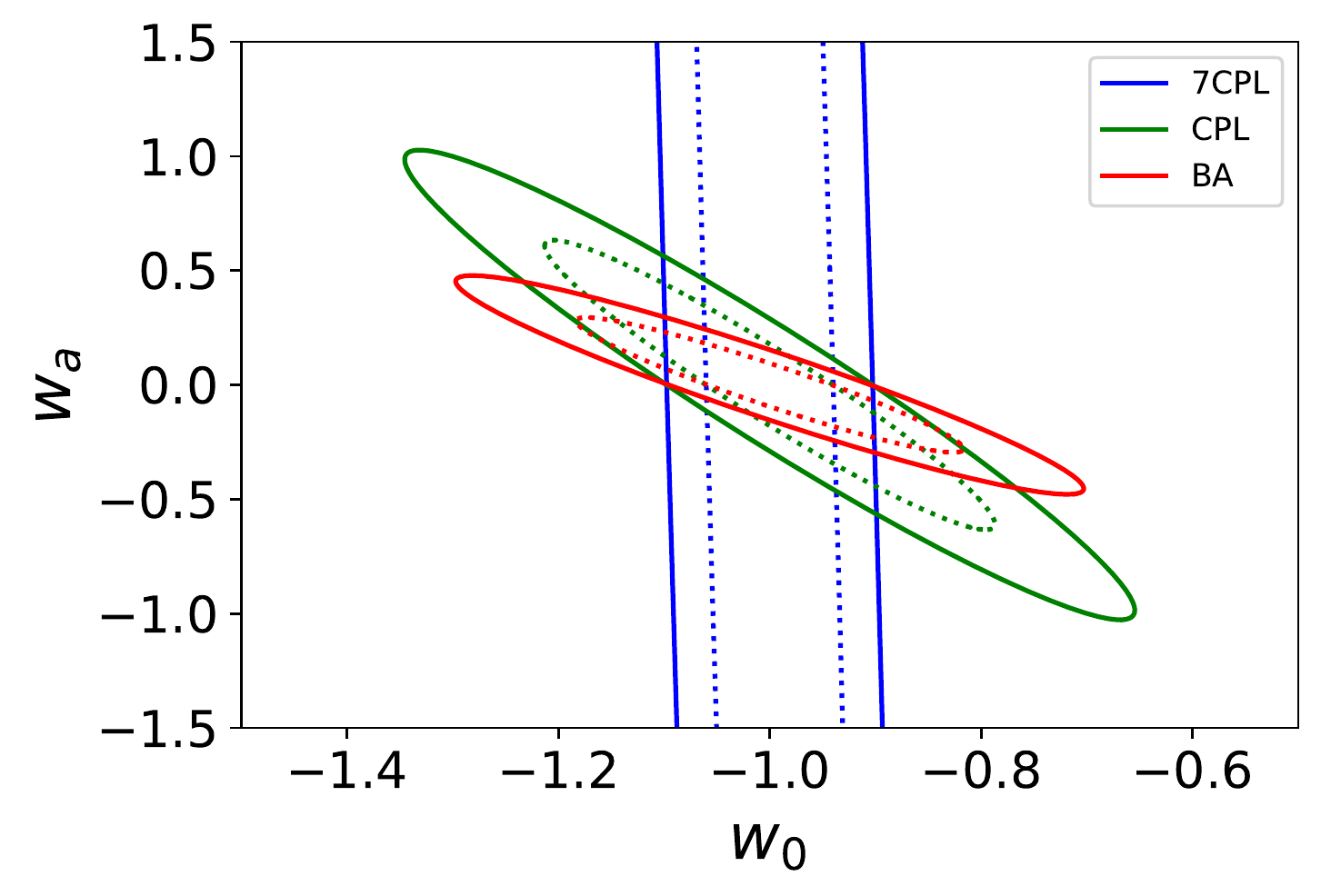}
\caption{ The $68\%$  and $95\%$ marginalized confidence intervals for the parameters $w_0$ and $w_a$ from the Lyman-$\alpha$ convergence cross correlation for different dark energy EoS parametrizations. The fiducial model is chosen as the LCDM model with $(w_0, w_a) = (-1,0)$. The three figures correspond to $\bar n_Q = 8 {\rm deg}^{-2}$, $\bar n_Q = 16 {\rm deg}^{-2}$ and $\bar n_Q = 100 {\rm deg}^{-2}$ respectively.}
\label{fig:Ly-contours}
\end{center}
\end{figure}
The Fisher matrix in Eq (\ref{eq:Fisher}) is used to put constraints on  the dark energy parameters.
We have assumed that other cosmological parameters are well constrained from CMBR observations and focus only on the dark energy EoS parameters.
Fig (\ref{fig:Ly-contours}) shows the  constraints on the parameters in the $w_0-w_a$ plane.
The contours correspond to  also show the $68\%$  and $95\%$ marginalized confidence intervals for the parameters $w_0$ and $w_a$.

\begin{table}
\caption{ The  $68 \%$ ($ 1 -\sigma$) Constraints of DE parameters  $(w_0, w_a)$ from Lyman-$\alpha$ and convergence cross power spectrum} 
\begin{center}
\begin{tabular}{c|c c c} 
 \hline 
 \hline 
  &  $\bar n_Q = 8 {\rm deg}^{-2}$  &  $\bar n_Q = 16 {\rm deg}^{-2}$ &  $\bar n_Q = 100 {\rm deg}^{-2}$ \\ 
 \hline
 BA& $\Delta w_0 = 0.17, \Delta w_a = 0.27 $&    $ \Delta w_0 = 0.15, \Delta w_a = 0.243 $  & $ \Delta w_0 = 0.141, \Delta w_a = 0.192 $\\ 
 CPL & $\Delta w_0 = 0.191, \Delta w_a = 0.57$ &   $\Delta w_0 = 0.168, \Delta w_a = 0.51$ & $\Delta w_0 = 0.121, \Delta w_a = 0.42$ \\ 
 7CPL &  $\Delta w_0 = 0.081 , \Delta w_a = --$   & $\Delta w_0 = 0.079, \Delta w_a = --$  & $\Delta w_0 = 0.068, \Delta w_a = --$ \\ 
 \hline
  \hline
    \end{tabular}
    \end{center}
    \label{table:sig}
     \end{table}
The table (\ref{table:sig}) shows the $1-\sigma$ errors on EoS parameters.
We find that if  dark energy is described by CPL parametrization, the constraint in  $w_0-w_a$  space is competitive with the  Planck+Bao+Supernova+HST for CPL parametrization \cite{refId0}. 

We also find that for the 7CPL parametrization, the parameter $w_a$  can not be constrained. However, for  7CPL parametrization we can 
constrain $w_0$ much better than CPL or BA. Thus we note that for the measurement of the present day value of the dark energy equation of state,
 7CPL is definitely favourable than CPL or BA and the  generally popular CPL parametrization may not be the best choice.
 It is also seen that the parameter space regions corresponding to $(w_0 > -1, w_a >0)$ and $ (w_0 < -1, w_a < 0)$ is constrained much better than 
 the regions corresponding to $(w_0 > -1, w_a < 0)$ and $ (w_0 < -1, w_a >  0)$.
 Thus regions which correspond to dark energy being modeled by only phantom or only non-phantom scalar fields are very strongly constrained.

\section{Cross-correlation of CMBR weak-lensing with redshifted 21-cm signal}
In a radio-interferometric observation the noise {\it rms}. in the real part in each visibility for single polarization measurement 
is given by 
\be
V_{rms}=\frac{T_{sys}}{K\sqrt{2 \Delta \nu \Delta t}} \frac{\lambda ^2}{2 k_B} 
\ee
Here $k_B$ is the Boltzmann
constant, 
  $\Delta \nu$ is the width of the frequency channels and $\Delta t$ is the integration time of the correlator.
  The total system temperature $T_{sys}$ maybe decomposed as 
$ T_{sys} = T_a + T_{sky} $, where  $T_a$ is an instrumental contribution and $T_{sky}$ is a sky contribution which is  usually subdominant and is given by \cite{Obuljen_2018}
\be T_{sky} = 60{\rm K} \left ( \frac{\nu}{300{\rm MHz}} \right )^{-2.55}.\ee 
The quantity $K$  denotes the antenna sensitivity and maybe written as 
$K={A_{eff}}/{2\,k_B}$ where  $A_{eff}$ denotes the  effective collecting  area of each
antenna. The factor $ \frac{\lambda ^2}{2 k_B}$ intensity to temperature units in the Raleigh Jeans limit.

The noise variance in the visibility correlation is written as \citep{poreion7}  
\be
\sigma^2_{VV}=\frac{8\,V_{rms}^4}{{\cal{N}}_p}
\ee
where ${\cal{N}}_p$ is the number of visibility pairs in a particular visibility bin $U$ to $U+ \Delta U$. This is given by 
\citep{poreion7, Bagla_2010}  
\be
{\cal N}_p=\frac{1}{2}\left[ \frac{N_{ant}(N_{ant} -1)}{2}\frac{T}{\Delta t}\Delta U^2
  \rho(U,\nu)\right]^2  \frac{2\pi U \Delta
  U}{\Delta U^2}
\ee
where $N_{ant}$ is the total number of antennas is the array, and  $T$ is the total observation time.
The normalized baseline distribution function $\rho (U, \nu)$ is given by a convolution of the  antenna distribution function with itself \cite{kkd_2007}.
\be
\rho(U, \nu) = \frac{c}{B} \int_0^{\infty} \rho_{ant}({\mathbf r}) \rho_{ant}({\mathbf r} - \lambda \mathbf U) \ee
where $B$ is the band width and  $c$ is fixed by the normalization $\int d^2{\mathbf U} \rho(\mathbf U) = 1$.

\begin{table}[ht]
\caption{ Specifications of the uGMRT like array used in the analysis } 
\centering 
\begin{tabular}{c c c c c c  } 
\hline\hline 
Observation time T (hrs)  &Freq. range(MHz) & $T_{sys}$(K) & $N_{ant}$  & $\Delta U$ & $A_{eff}$($m^2$) \\ [0.5ex] 
\hline
 400 & 550-850  & 70 & 60, 100 & 32  & 1590  \\ [1ex]
\hline 
\end{tabular}
\label{tab:Telescope}
\end{table}

We consider Planck sensitivies for the cross-correlation. 
For the redshifted 21-cm observation we first consider  a futuristic version of the upgraded GMRT (uGMRT) \cite{2017CSci..113..707G}, with a larger number of antennas. The  antenna distribution is roughly assumed to behave as $ \rho_{ant}(r) \propto  1/r^2$  distributed over a $2 \times 2{\rm Km}^2$ region. Each antenna is assumed to be of $45$ m in diameter, with a
field of view of $1.3^{\circ}$ (FWHM).
 The frequency separation $\nu$ over which the 21-cm signal remains correlated  scales approximately as 
$\nu = 1 {\rm MHz} (\ell/100)^{-0.7}$ \cite{poreion7}. We assume
that the signal is averaged over frequency bins of this width
to increase the signal-to-noise ratio (S/N). 
 The different  telescope parameters used for our analysis is summarized in table (\ref{tab:Telescope}).

\begin{figure}[h]
\begin{center}
\includegraphics[height=6cm]{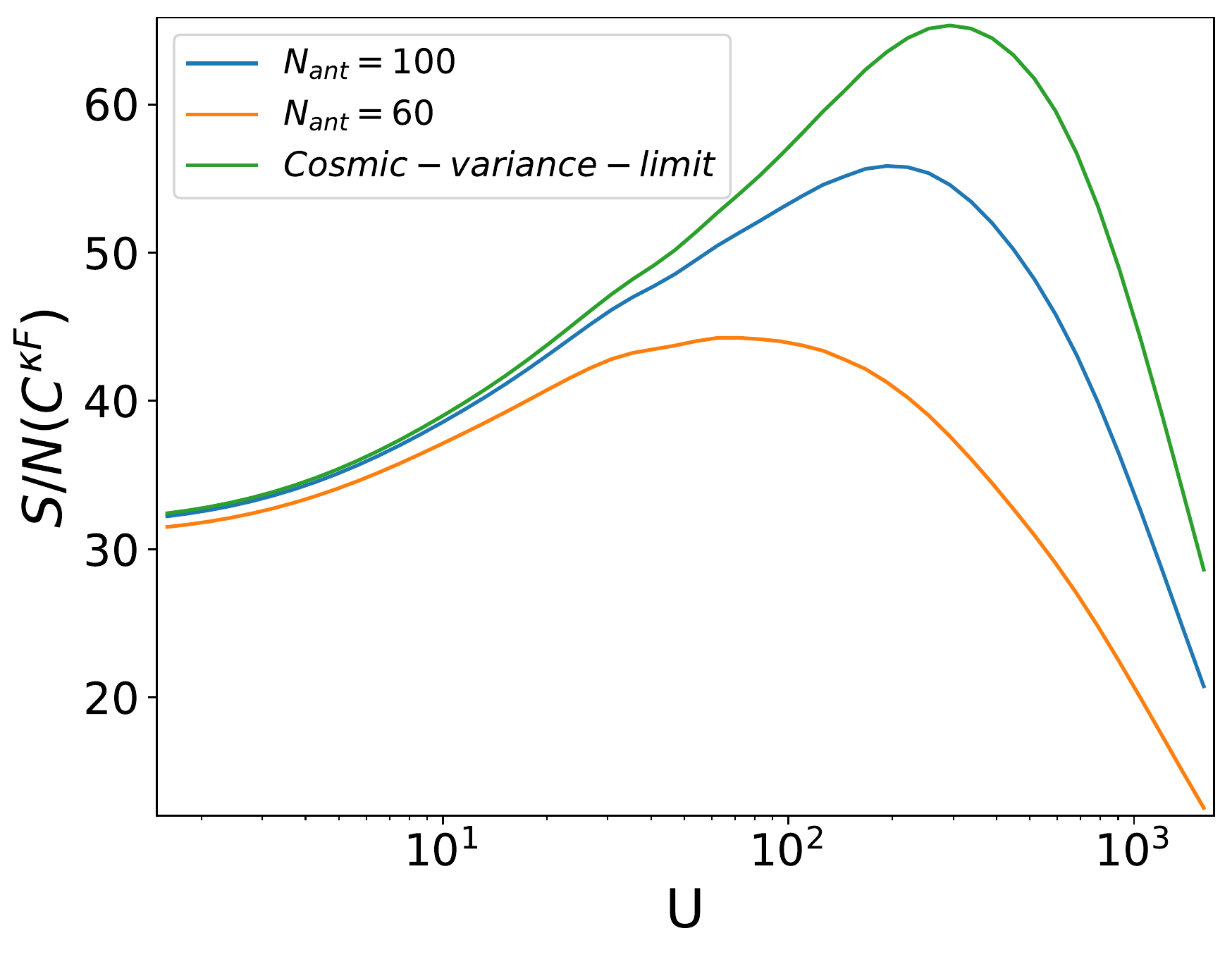}
\caption{ Signal to Noise ratio for the Convergence-21 cm cross angular power spectrum for the fiducial LCDM model for uGMRT like array  with  $N_{ant} = 60, 100 $ respectively. We also show the cosmic variance limit.}
\label{fig:SNR21}
\end{center}
\end{figure}
  
Figure (\ref{fig:SNR21}) shows the SNR for a LCDM fiducial model.
We find that for a uGMRT like array with $N_{ant} = 60$ with LCDM fiducial model a peak SNR of  $\sim 40$  is obtained at $ \ell \sim 600$. If the number of antennas in the interferometer is $N_{ant} = 100$ a peak SNR of $\sim 54$ is achieved at $ \ell \sim 1500$.
In the limit of negligible instrumental noise a peak SNR of $70$ is possible.

\begin{figure}[h]
\begin{center}
\includegraphics[height=6cm]{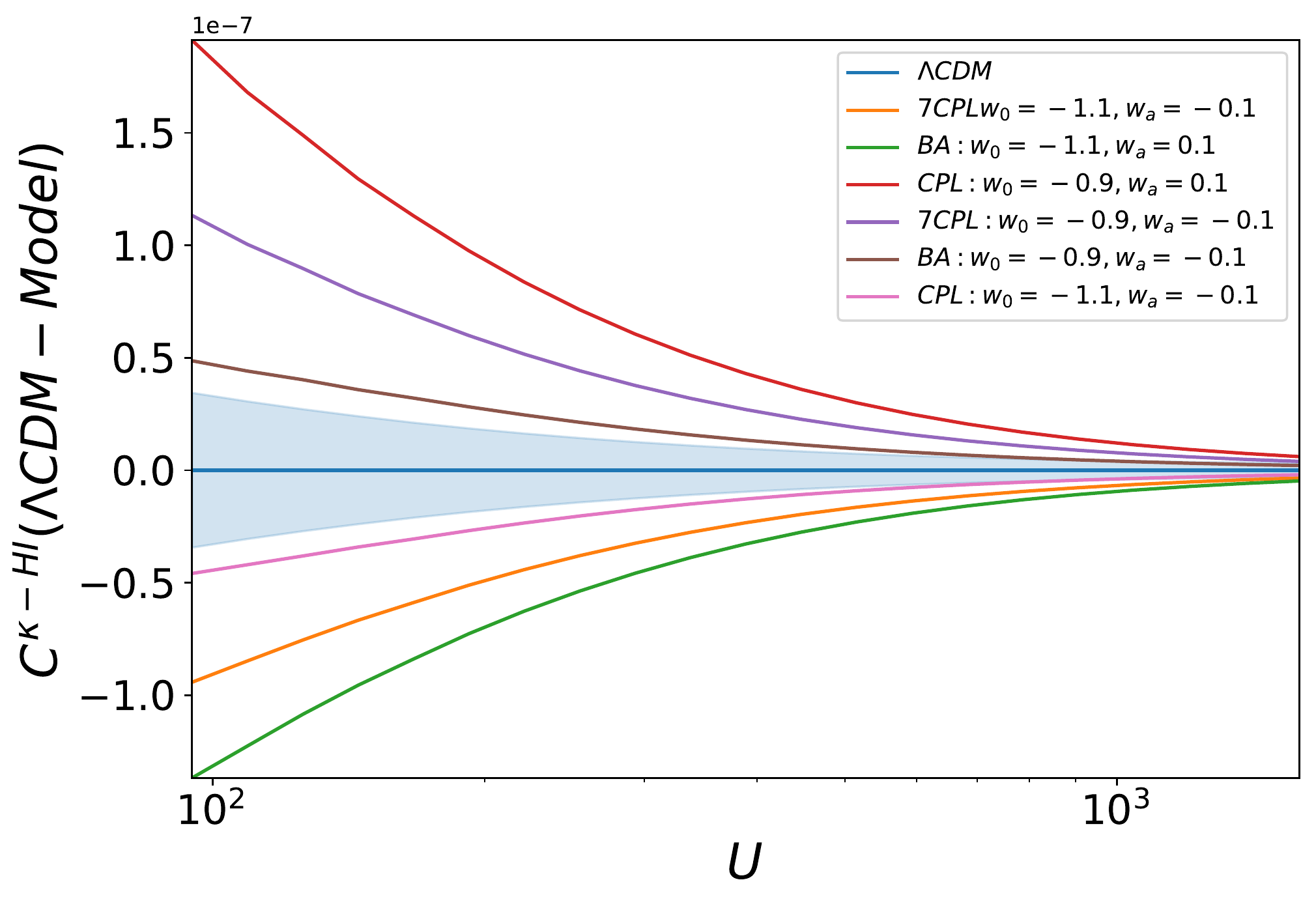}
\includegraphics[height=6cm]{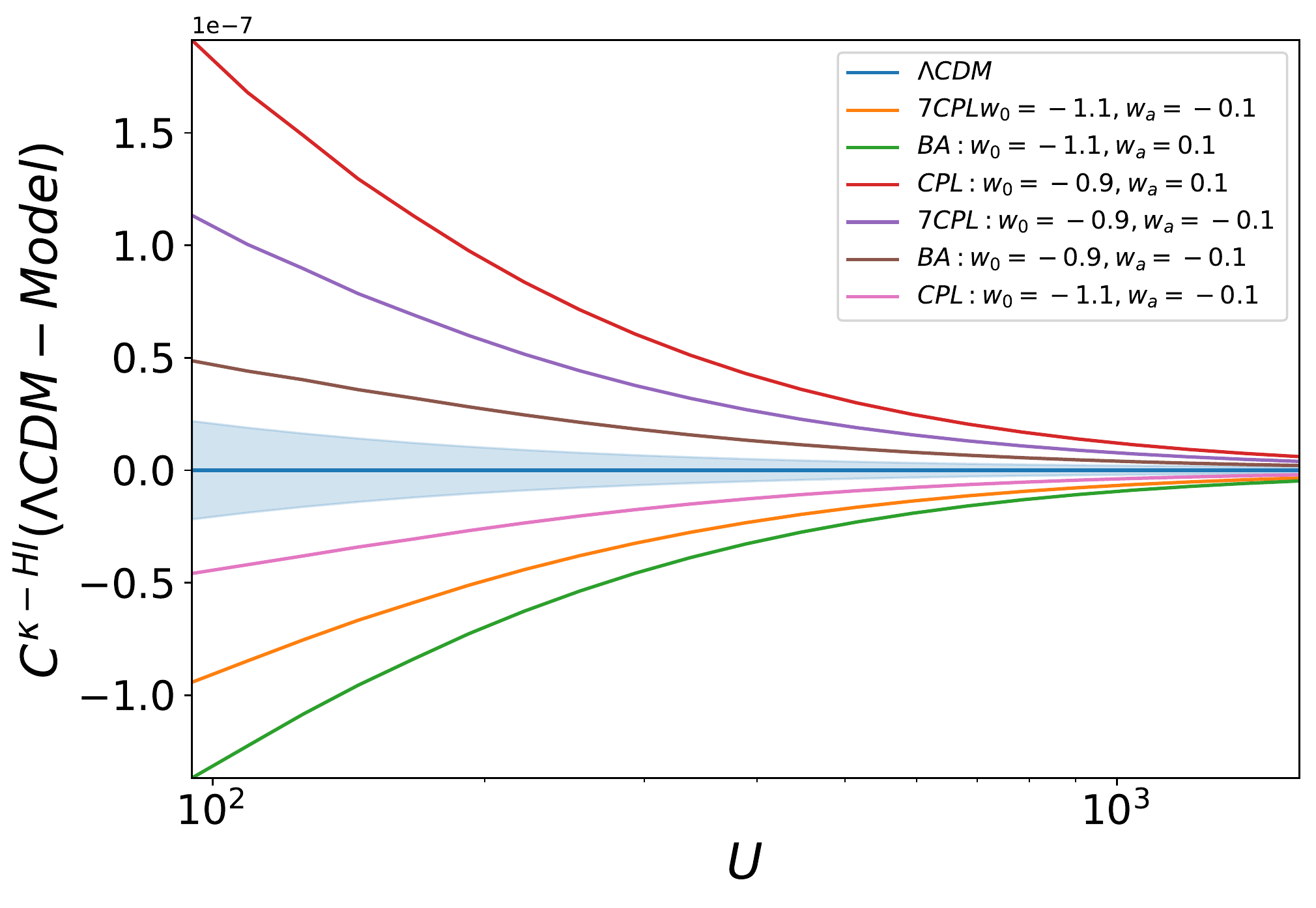}
\includegraphics[height=6cm]{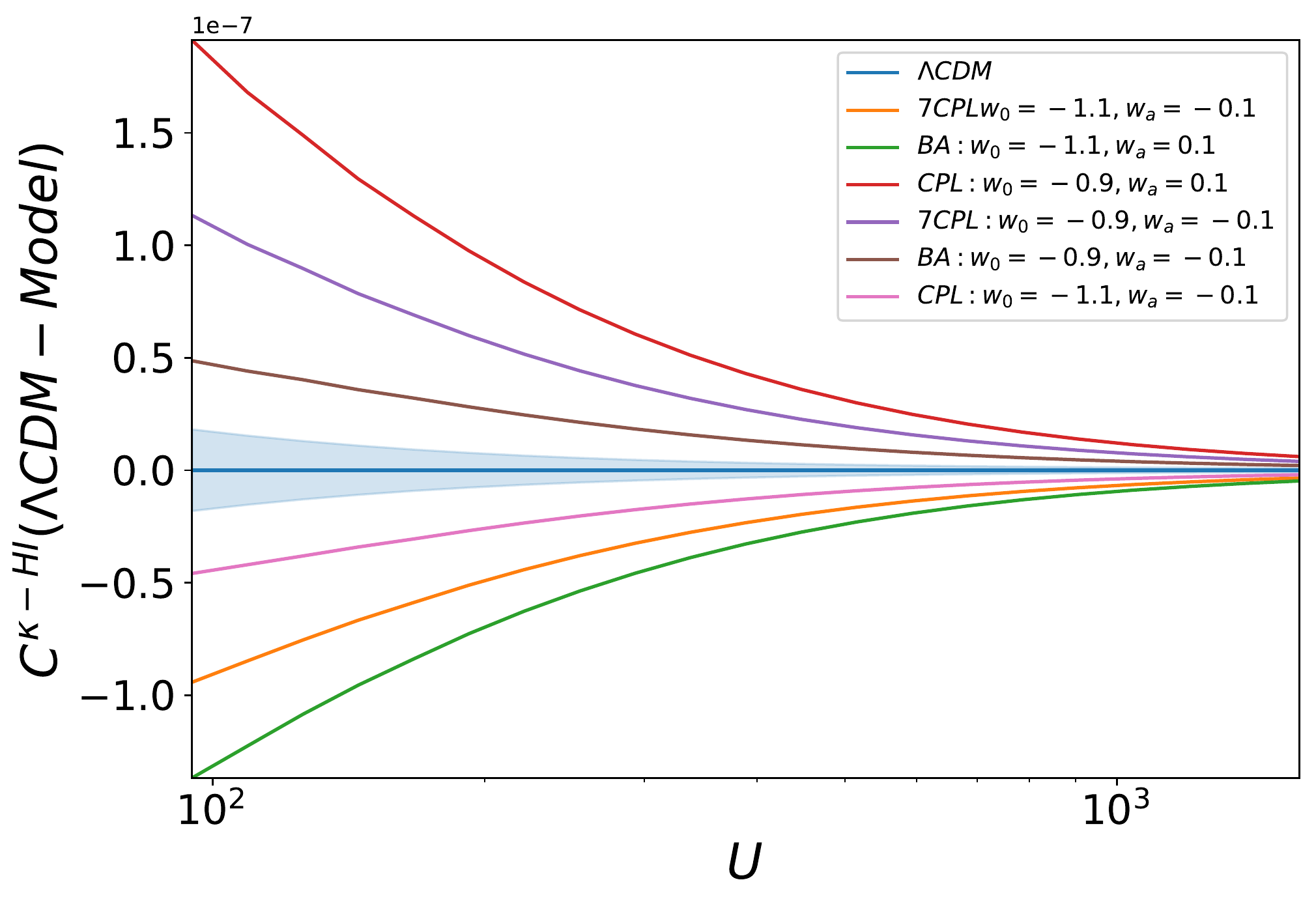}
\caption{ The  difference of the 21 cm  - convergence cross power spectrum for different fiducial dark energy parametrizations
from the  the $\Lambda CDM$ model. The $1-\sigma$ error is shown by the shaded region. The first and the second figure corresponds to a uGMRT like radio interferometer with $N_{ant} = 60 $ and   $N_{ant} = 100 $ respectively. The observation time is $400$ hrs. The last figure shows the best case scenario in the cosmic variance limit of no observational noise. }
\label{fig:signal-cross-21}
\end{center}
\end{figure}

Figure (\ref {fig:signal-cross-21}) shows the difference of the  21-cm and CMBR convergence cross angular power spectrum 
for different dark energy EoS parametrizations from the cross angular power spectrum for the LCDM model.
We find that BA,  CPL and 7CPL models to be differentiable from the LCDM model at $ > 3 \sigma $ sensitivity.

For making  more realistic  error projections on $( w_0-w_a)$,  we also consider the radio interferometer MeerKAT \cite{inproceedings}.
This telescope has dishes of $13.5$m   diameter and a wider field of view.  We consider the UHF-band of this telesope for our analysis.
The telescope parameters for MeerKAT are summarized in the table (\ref{tab:Telescope1}). 
\begin{table}[ht]
\caption{MeerKAT specifications} 
\centering 
\begin{tabular}{c c c c c c } 
\hline\hline 
Observation time T (hrs)  &Freq. range(MHz) &  $T_{sys}$ (K) & $N_{ant}$&   $\Delta U$ & $A_{eff}/ T_{sys}$($m^2/K$) \\ [0.5ex] 
\hline
400 & 580-1015 &  30 & 64 & 12 &  320  \\ [1ex]
\hline 
\end{tabular}
\label{tab:Telescope1}
\end{table}

\begin{figure}[h]
\begin{center}
\includegraphics[height=5.0cm]{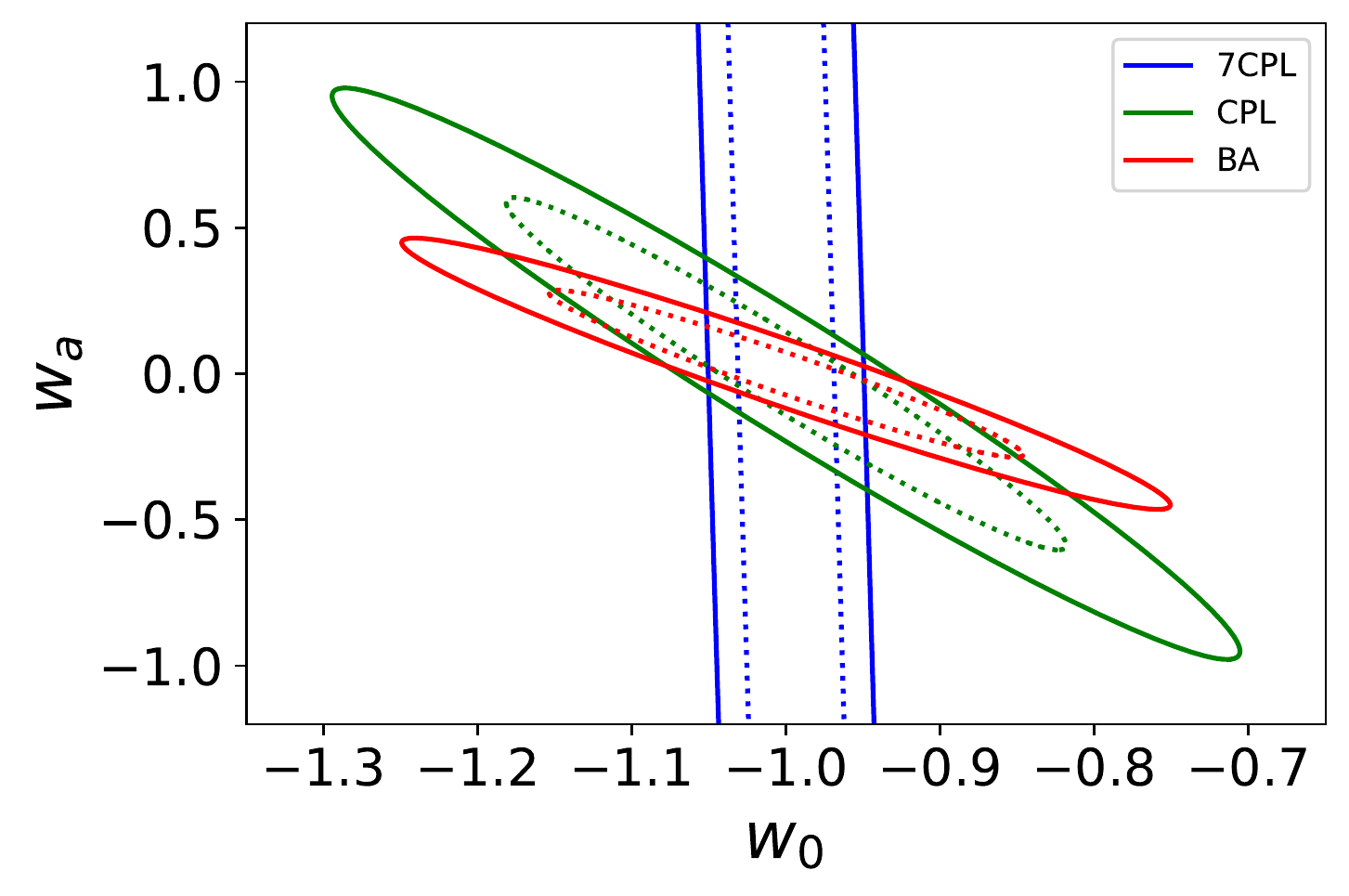}
\includegraphics[height=5.0cm]{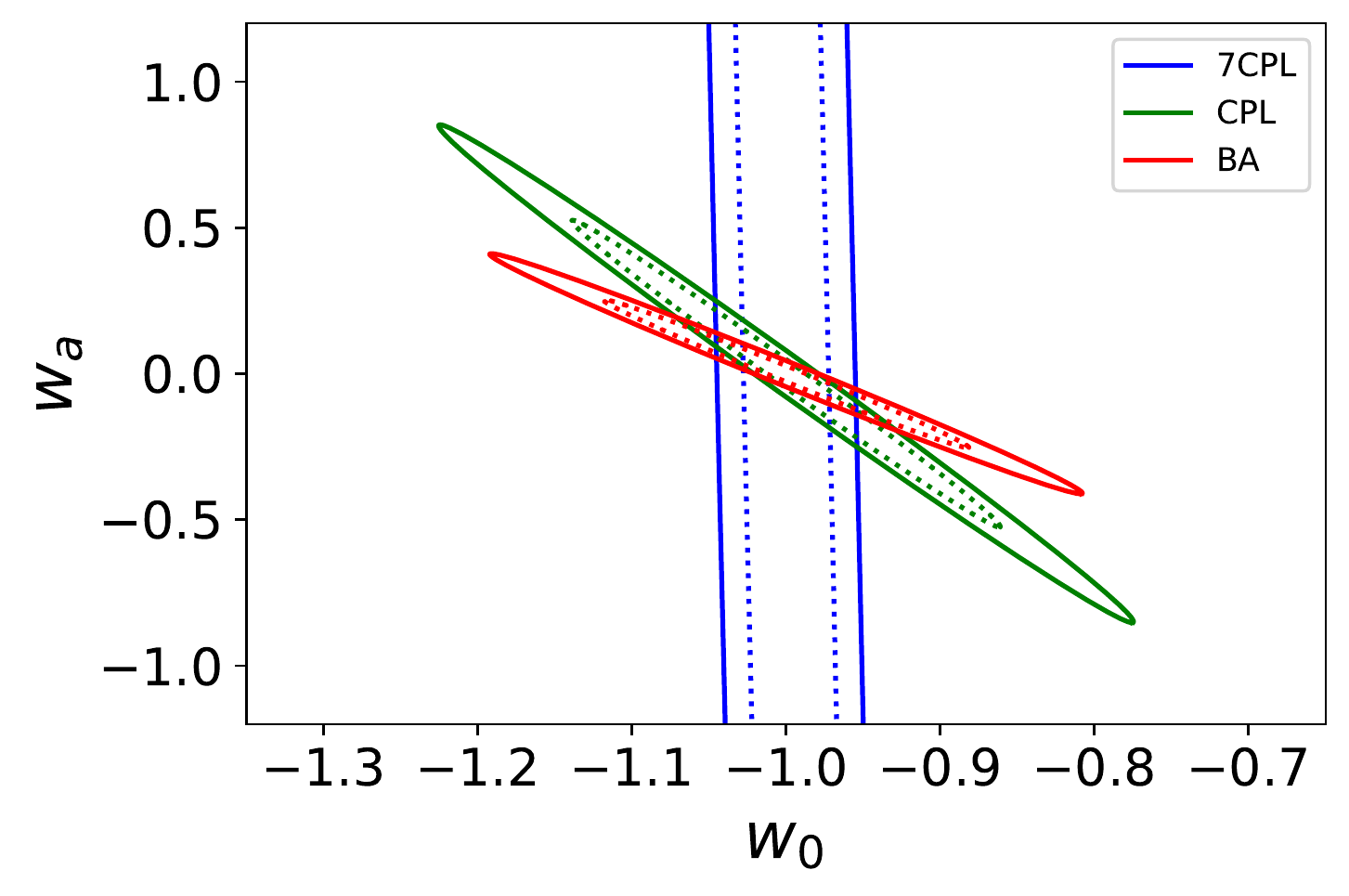}
\caption{ The $68\%$  and $95\%$ marginalized confidence intervals for the parameters $w_0$ and $w_a$ from the 21-cm convergence cross correlation for different dark energy EoS parametrizations for uGMRT like telescope with $N_{ant} = 100$. The fiducial model is chosen as the LCDM model with $(w_0, w_a) = (-1,0)$.  The second figure shows the best case scenario of a cosmic variance dominated experiment.}
\label{fig:21-contours-ugmrt}
\end{center}
\end{figure}

\begin{figure}[h]
\begin{center}
\includegraphics[height=5.252cm]{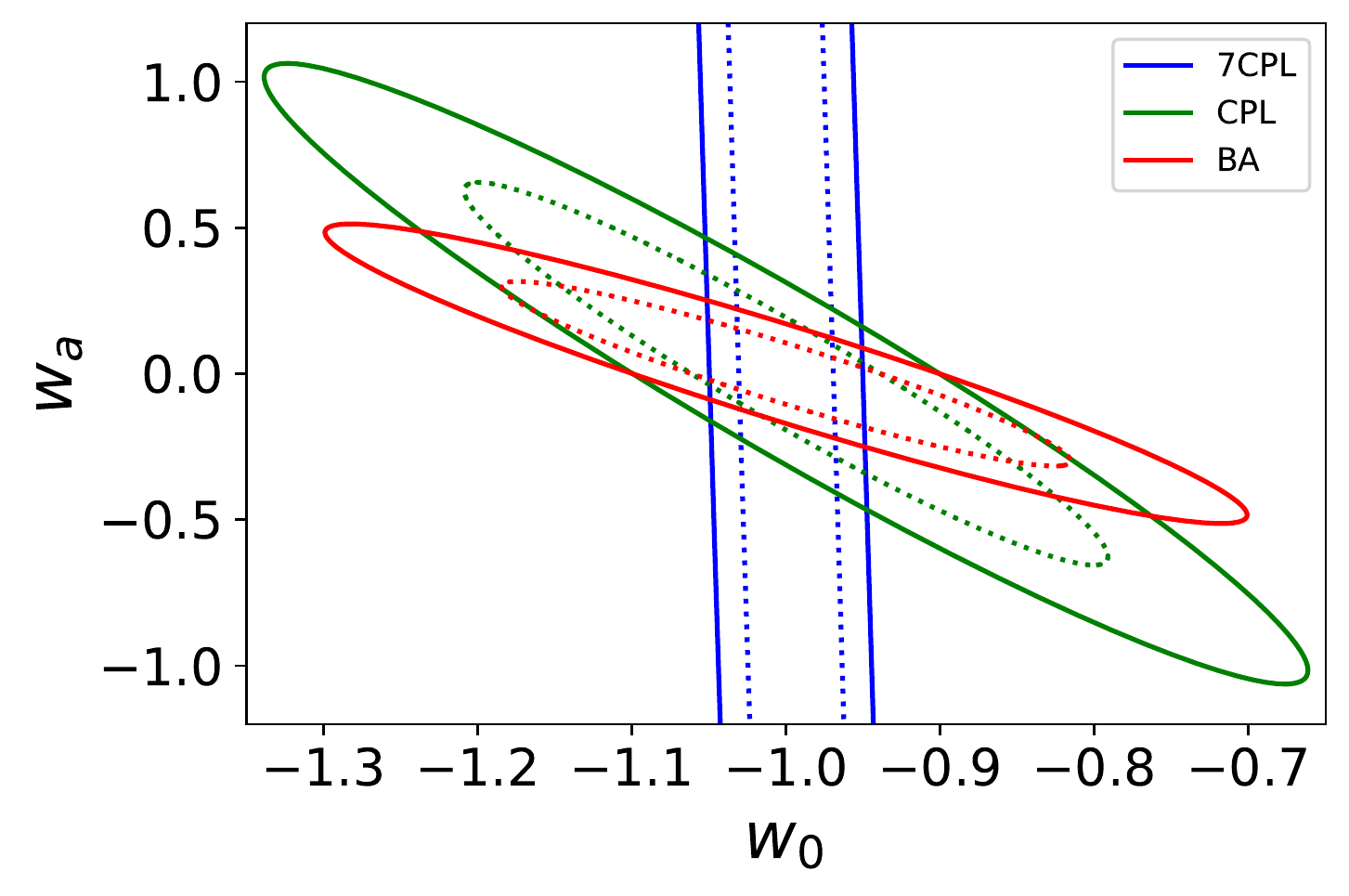}

\caption{ The $68\%$  and $95\%$ marginalized confidence intervals for the parameters $w_0$ and $w_a$ from the 21-cm convergence cross correlation for different dark energy EoS parametrizations. The fiducial model is chosen as the LCDM model with $(w_0, w_a) = (-1,0)$. The figures correspond to MeerKAT.}
\label{fig:21-contours-meerkat}
\end{center}
\end{figure}

The Fisher matrix in Eq (\ref{eq:Fisher}) is used to make error projections for the dark energy parameters.
Fig (\ref{fig:21-contours-ugmrt}) shows the  constraints on the parameters in the $w_0-w_a$ plane for cross-correlation with a  uGMRT like telescope.
The contours correspond to  the $68\%$ and $95\%$ marginalized confidence intervals for the parameters $w_0$ and $w_a$.
A summary of the errors obtained in the parameters $(w_0, w_a)$ is given in the table(\ref{tab:param}). We find that the constraints on the EoS parameters are much better than the ones obtained from the cross-correlation with Lyman-$\alpha$ forest. This is due to the large bandwidth of the 21 cm observation over which the signal is averaged.

The results from cross-correlation with a more realistic MeerKAT is shown in the figure (\ref {fig:21-contours-meerkat}). 
The marginalized 1-$\sigma$ errors obtained in the parameters $(w_0, w_a)$ for MeerKAT  is also given in the table(\ref{tab:param}). The constraints show reasonable degradation as compared to the   uGMRT predictions (with $N_{ant} = 100$).

\begin{table}[h]
\caption{The  $68 \%$ ($ 1 -\sigma$) Constraints of DE parameters  $(w_0, w_a)$ from HI 21-cm and convergence cross power spectrum}
\centering
\begin{tabular}{c c c c }
\hline 
\hline
Model & Cosmic variance limit  &  uGMRT($N_{ant} = 100$) & MeerKAT \\ \hline
BA  & $\Delta w_0 = 0.079$, $\Delta w_a =  0.121$  & $\Delta w_0 = 0.098$, $\Delta w_a =  0.20$ & $\Delta w_0 =0.122 $, $\Delta w_a =  0.22$ \\ 
CPL & $\Delta w_0 =  0.092$, $\Delta w_a = 0.168$  & $\Delta w_0 = 0.115$, $\Delta w_a =   0.40$ & $\Delta w_0 = 0.138$, $\Delta w_a =  0.434$ \\ 
7CPL & $\Delta w_0 = 0.04$, $\Delta w_a = -- $  & $\Delta w_0 = 0.05$, $\Delta w_a =   --$ & $\Delta w_0 = 0.053$, $\Delta w_a =  --$ \\
\hline \hline
\end{tabular}
\label{tab:param}
\end{table}

 The projected constraints on the dark energy equation of state  indicate that for a broad band 21-cm survey with uGMRT like telescope with 100 or more antennas the cross-correlation may constrain dark energy evolution at a level which is competitive with the  projections for auto-correlation results with a SKA1 type experiment \cite{dinda2018} or a  joint Planck+SN+BAO+HST \cite{refId0}. This is partly due to the fact that the projected  instrumental noise levels with futuristic telescopes are not very high. However, we emphasize that in our analysis of cross-correlation of the tomographic field of 21-cm signal with the integrated weak lensing field, we have abandoned the line of sight information. By averaging the 21-cm signal along the radial direction, the SNR  has improved considerably as compared to the ususal tomographic 3-dimensional predictions from the auto-correlation.

The 7CPL parametrization is seen to constrain the present value of dark energy EoS much better than CPL and BA parametrization which is also the conclusion drawn in earlier works \cite{dinda2018}. 
We also find that BA model constrains $(w_0, w_a)$ much better than CPL model. Since CPL parametrization is the widely use parametrization to model dark energy evolution, we note that it may not be suitable to constrain dark energy evolution. 
The projection of ($w_0,w_a$) for CPL model has been studied extensively using eBOSS and DESI cosmological data \cite{Zhao_2016,vargasmagana2019unraveling}. 
The limits of the  error contours corresponding to the CPL model are $w_0 = ( -1.5, -0.5)$, $w_a= (-1, 1)$ for the eBOSS survey which is similar to our projections. The   PLANCK+DESI data  limits for $w_0$ and $w_a$ are  $w_0 =( -1.13, -0.86)$  and $w_a= (-0.4, 0.4)$ respectively, which has better figure of merit.
A comparison of dark energy EoS parameter error projections for different models are studied in the paper using Sne Ia JLA and BAO datasets \cite{Escamilla_Rivera_2016}. Their projected error limits for the CPL model parameters are $ \approx w_0 = (-1.6, -0.75)$ and  $w_a = (-2.5, -1.5)$. 
Thus, we note that the cross-correlations between the weak lensing convergence with HI-21cm and Ly-$\alpha$ forest found better or reasonably close constraints on the $w_0,w_a$ parameter space when compared with some other probes.

\section{Conclusion}

There are several observational aspects that we have not considered in this work.
Foreground subtraction issues, is a  major concern for the 21-cm signal. Large astrophysical foregrounds from galactic and extra
galactic sources pose a serious threat towards achieving desired detection sensitivities \cite{Ghosh_2010}. 
Though the problem of foreground subtraction is less for the cross correlation, a significant amount of foreground subtraction is
required to obtain good SNR since the foregrounds appear as noise in the
cross-correlation. For the Lyman-$\alpha$ forest observations, continuum subtraction and avoiding 
metal line contamination, though less serious, needs to carried out with high precision.
Further, man made radio frequency interferences (RFIs), calibration errors and
other systematics also needs to be tackled for a detection of the HI 21-cm signal. 
We emphasize that some of the fundamental problems posed by 21-cm foregrounds can be avoided by considering cross correlations.

 We also note that we have restricted our study of dark energy evolution to a very  small set of commonly used parametrizations. There are a large number of possible 2-parameter descriptions which may suit some specific dark energy dynamics \cite{2016PhRvD..93j3503P}.  Though the CPL model and its variants fits a reasonable stretch of dark  energy dynamics they do  not describe all kinds of dynamics \cite{2008GReGr..40..329L, PhysRevLett.92.241302}. We also note that the error projections obtained in this paper from a Fisher matrix analysis gives a good idea about the efficacy of the cross-correlation in constraining the $(w_0, w_a)$ parameter space. However, a more sophisticated Bayesian analysis is needed for more robust statistical predictions.

We conclude by noting that the cross-correlation power spectrum of the HI tracers from the post
reionization epoch with weak-lensing fields is a  direct probe of cosmological structure formation which can be detected to a
high level of statistical sensitivity with upcoming QSO surveys and radiointerferometers.
This has the potential to provide new insights on  dark energy evolution. This probe may be combined  with other cosmological observations like CMB, BAO, SNIa etc and a joint analysis  shall be able to give us a clearer picture of the nature of dark energy and its cosmic evolution.
\bibliography{references}{}
\bibliographystyle{JHEP}
\begin{centering}
\section*{Appendix}
\end{centering}
\subsection*{Background Evolution}

\begin{figure}[h]
\begin{center}
\includegraphics[height=5cm]{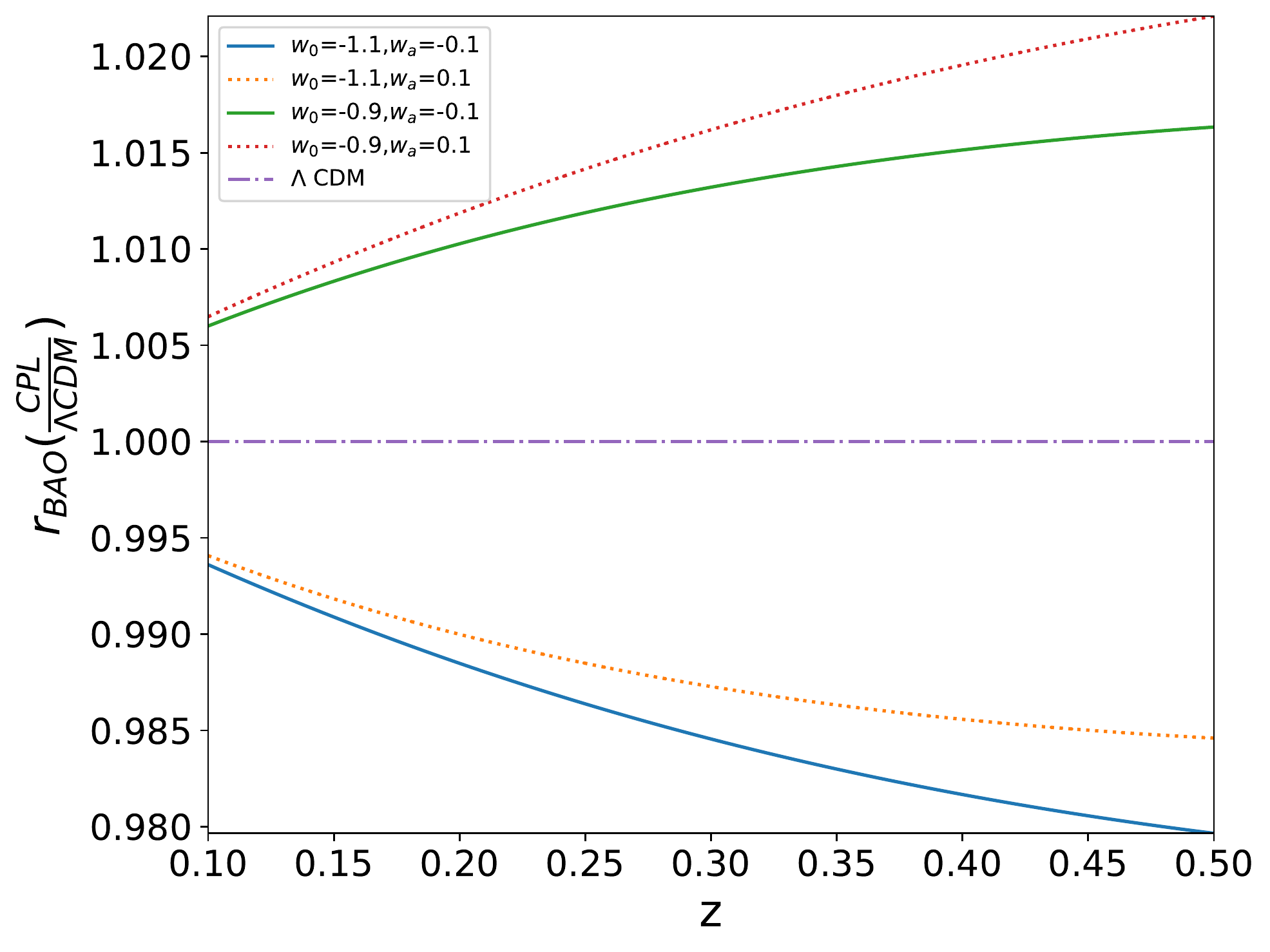}
\includegraphics[height=5cm]{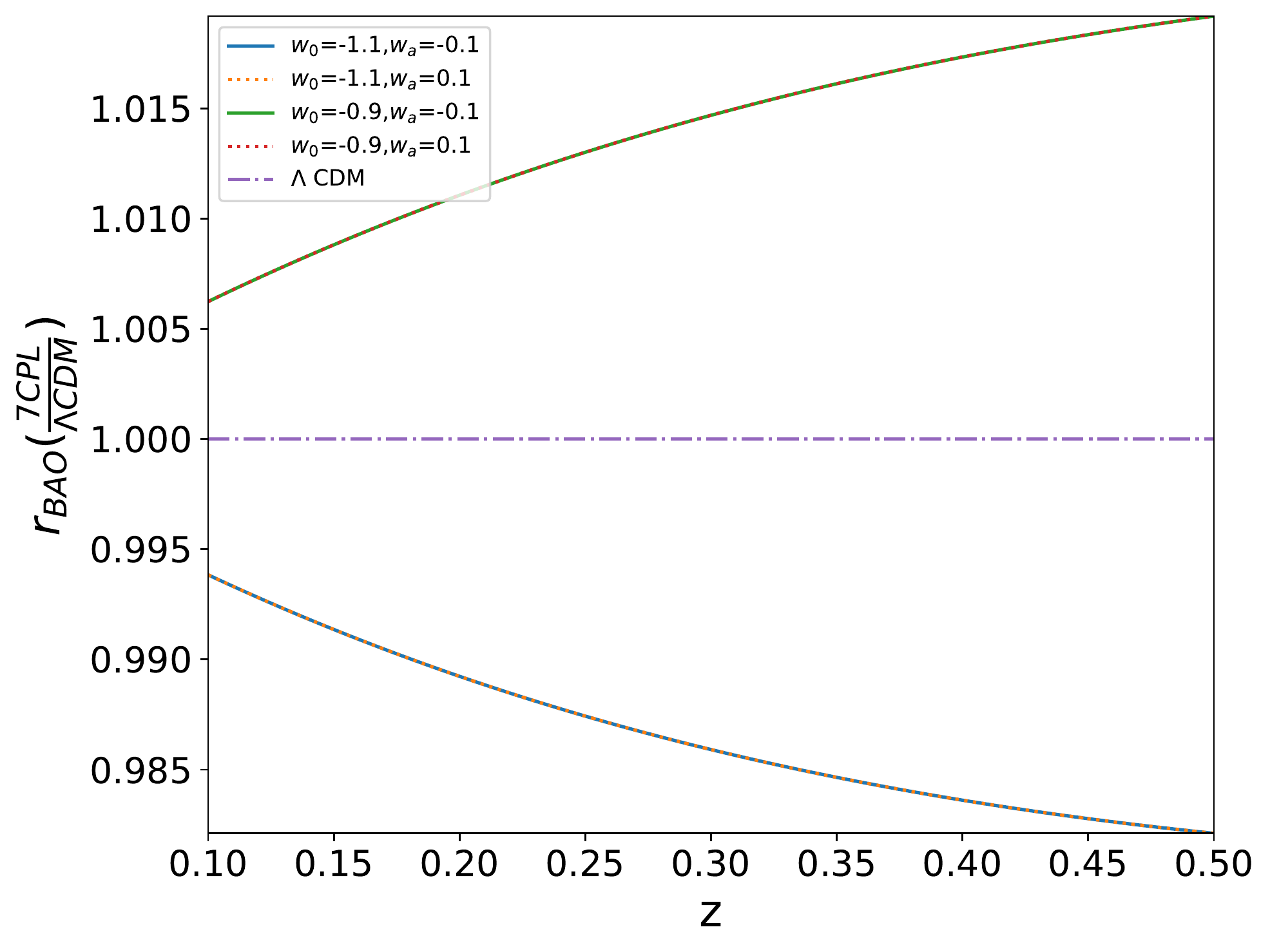}
\includegraphics[height=5cm]{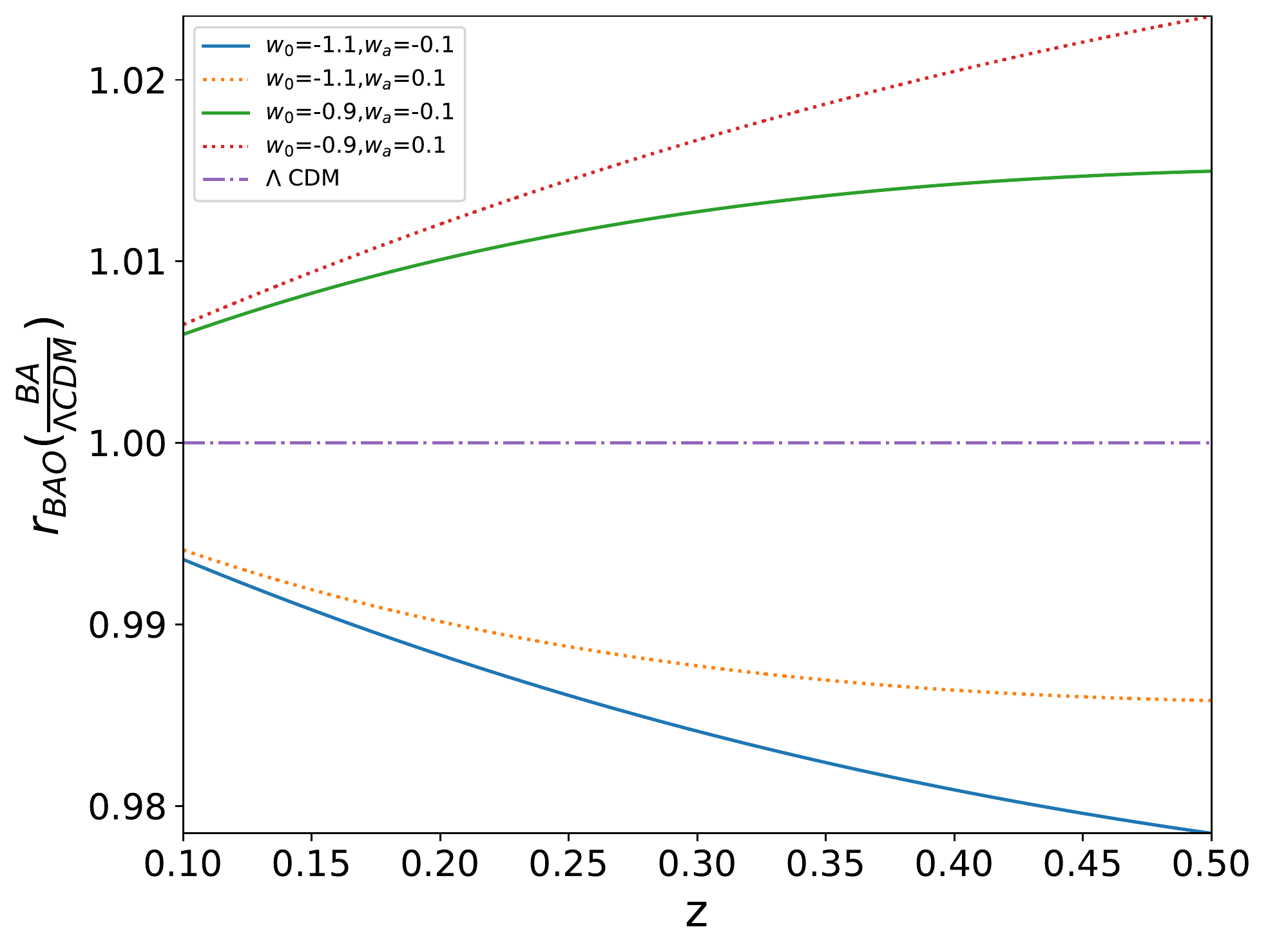}

\caption{ The BAO distance ratio  $r_{_{BAO}} (z)$ versus redshift $z$ for different dark energy models (CPL, 7CPL, BA) in a spatially flat Universe scaled by $r_{_{BAO}} (z)^{\Lambda CDM}$ for the $\Lambda CDM$ model. }
\label{fig:bao}
\end{center}
\end{figure}
\begin{figure}[h]
\begin{center}
\includegraphics[height= 5cm]{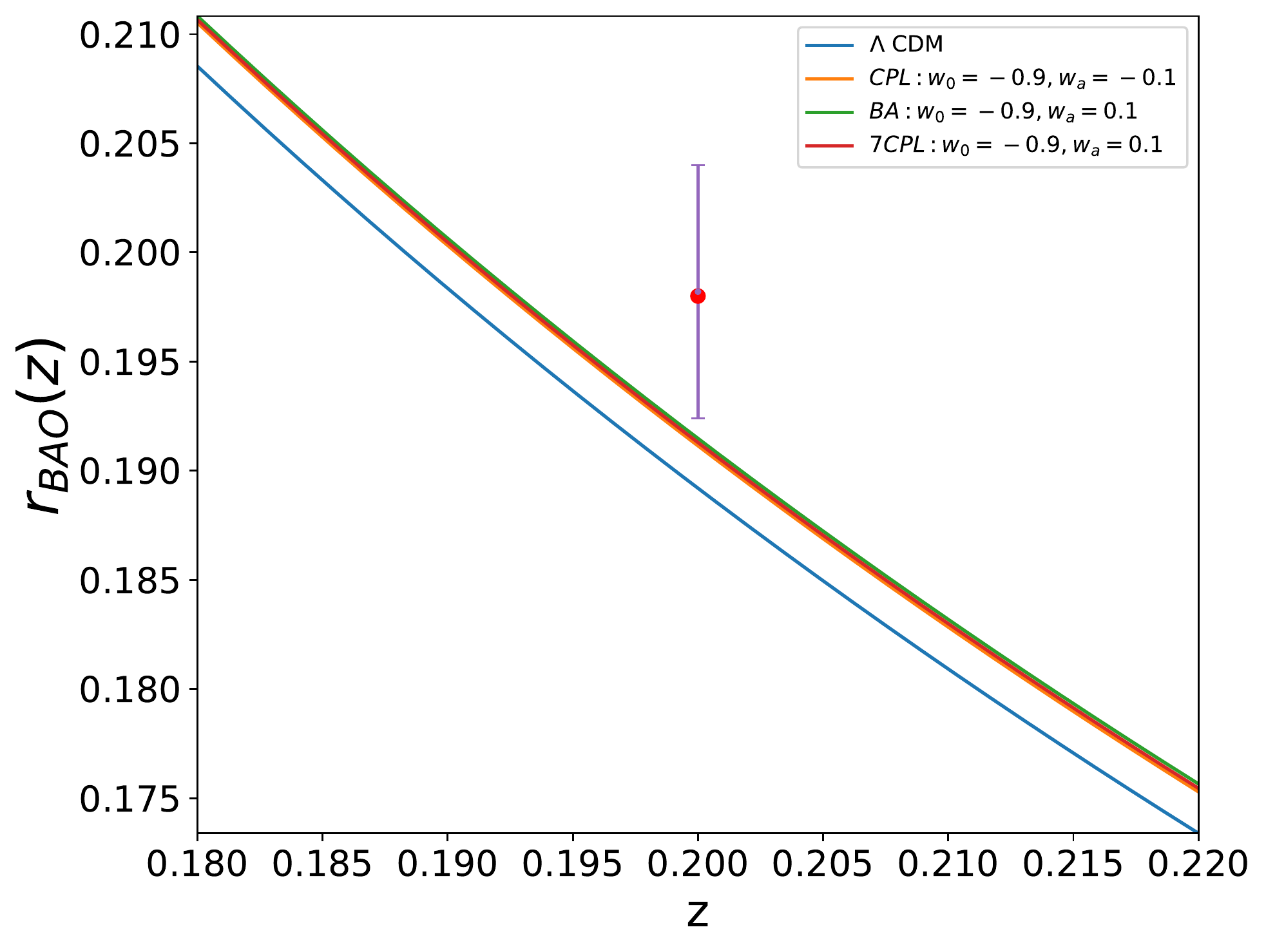}
\includegraphics[height=5cm]{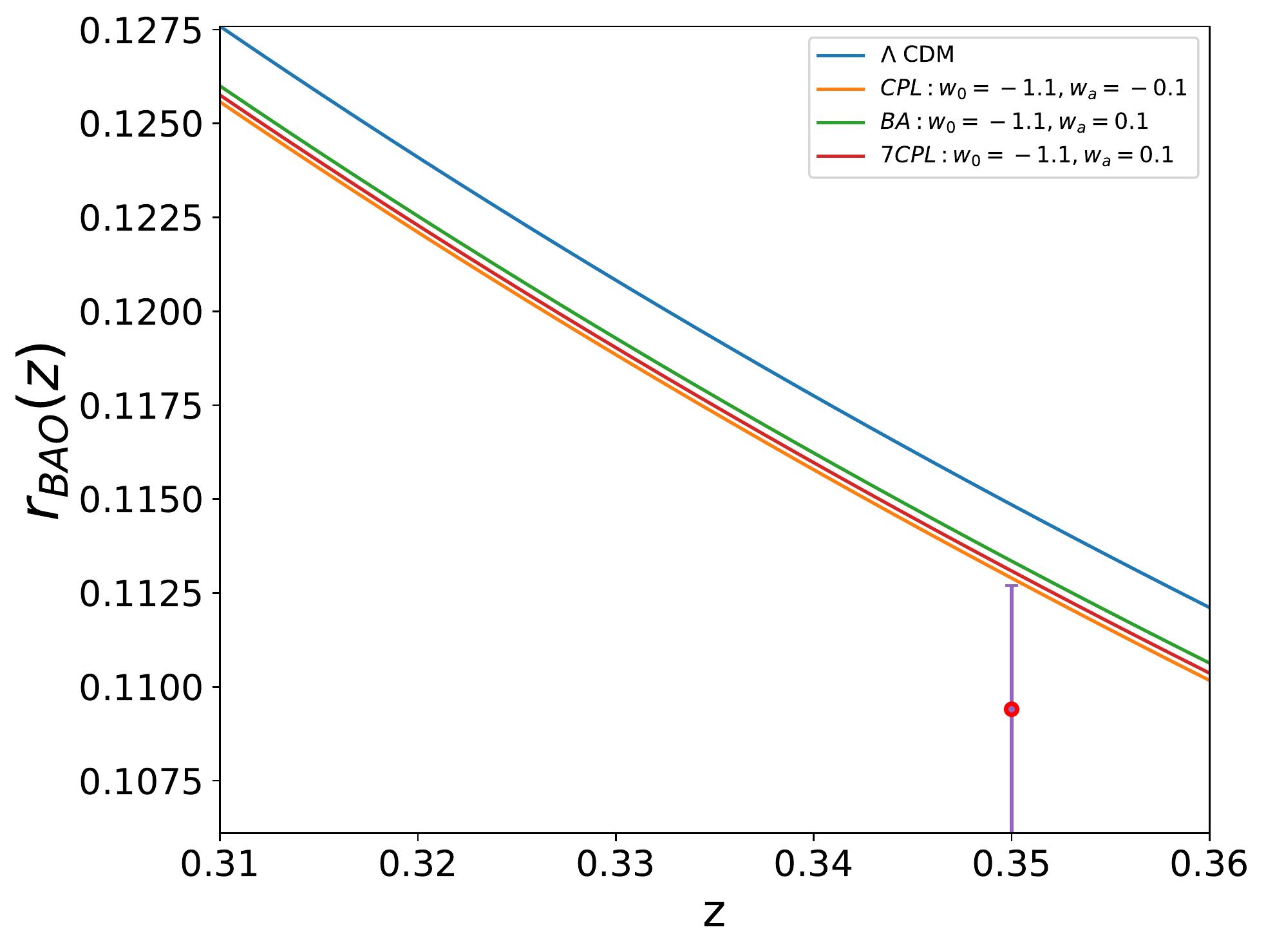}
\includegraphics[height=5cm]{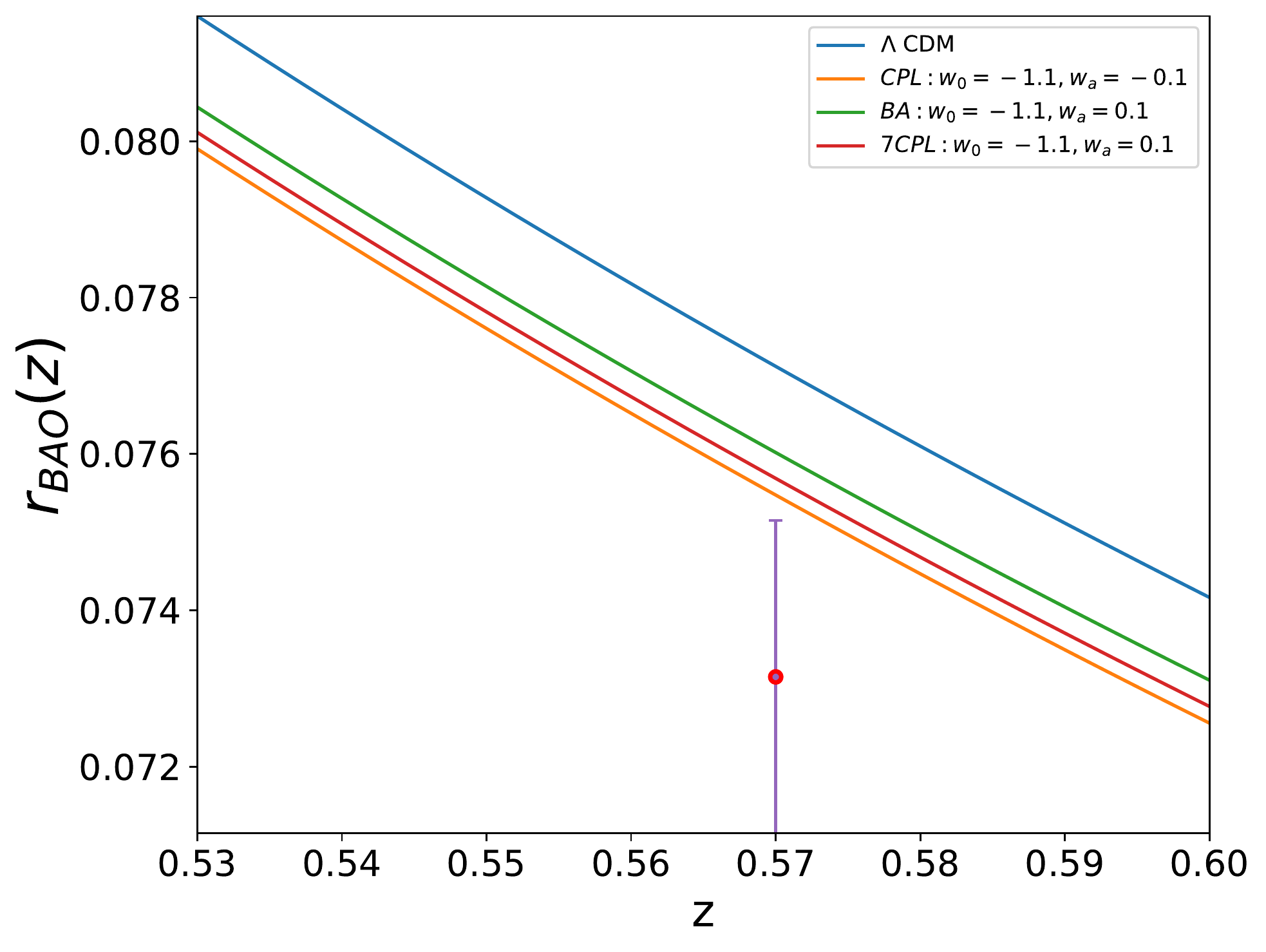}

\caption{ The first two figures in the panel shows the data points from the 2df galaxy survey at redshifts $z = 0.2$ and $ z= 0.35$ respectively and the third figure shows the high redshift data  at 
$z = 0.57$ from BOSS SDSS-III survey. }
\label{fig:baodata} \end{center}
\end{figure}

Baryon acoustic oscillation (BAO) observations \cite{White_2005}  aim to constrain the angular diameter distance $d_{A}(z)$ and the Hubble parameter $H(z)$ 
through the imprint of the oscillatory feature of  the matter power spectrum in the transverse (angular) and longitudinal directions respectively.
Due to low SNR in BAO measurements, it is often convenient to measure an effective distance defined as \citep{amendola_tsujikawa_2010}
\be
D_V (z) = \left [(1 + z)^2 d_A(z)^2  \frac{c z}{H(z)} \right ]^{1/3} \ee
This effective distance  is a direct quantifier of the background cosmological model (density parameters) and is thereby sensitive to the dynamical evolution of  dark energy.
We use a dimensionless quantifier of cosmological distances \cite{Eisenstein_2005}
\be
r_{_{BAO}} (z) = \frac{r_s}{D_V(z)} \ee
where $r_s$ denotes the sound horizon at the recombination epoch.
Figure (\ref{fig:bao}) shows the departure of $r_{_{BAO}}$ as a function of $z$ from the $\Lambda CDM$ model prediction. A redshift dependent difference of a few percent from the   $\Lambda CDM$ model is seen for the different parametrizations. The behaviour is very similar for the  CPL and BA parametrizations which are known to mimic the thawing class of dark energy models. The 7CPL parametrization which represents the tracking class shows identical behaviour for some sets of  model parameters (see figure (\ref{fig:bao})).
Observations from the 2df galaxy redshift survey gives the bounds on $r_{_{BAO}}$ as
$r_{_{BAO}}( z = 0.2) = 0.1980 \pm 0.0058$ and $r_{_{BAO}}(z = 0.35) = 0.1094 \pm 0.0033$ \citep{Percival_2007}.  All the models with a redshift dependent $w(z)$ seems to be in better agreement with this data.
The analysis of BOSS (SDSS III) CMASS sample along with Luminous red galaxy sample \cite{Anderson_2012}  from SDSS-II  gives  $r_{_{BAO}}( z = 0.57) = 0.07315 \pm 0.002$. 
Figure (\ref{fig:baodata}) shows the model predictions as compared to the data at different redshifts.

\subsection*{Growth of Perturbations}
At sufficiently early times and on large spatial scales the matter density fluctuations  are much less than unity ($|\delta|<<1$ , in linear regime).  Under these conditions the  matter density contrast $\delta = \frac{\delta\rho_m} {\bar \rho_m}$ evolves independently for different Fourier modes $k$. Separating the time dependent part of the matter overdensity as $\delta{(a)}= D_+(a)\delta_m(a=1)$. The growth of density fluctuations on sub horizon scales can be obtained by solving the second order differential equation 
\begin{equation}
\frac{d^2 D_+}{da^2} +  \left ( \frac{1}{H}  \frac{dH}{da}  +  \frac{3}{a} \right ) \frac{dD_+}{da} -  \frac{3}{2}\frac{\Omega_{m_{0}}  H_0^2}{a^5 H^2 }  D_+ = 0
\label{eq:growingmode}
\end{equation}
In order to solve the above equation we choose the initial conditions in the matter dominated epoch ($a_i=10^{-3}$), where growth function grows linearly with the scale factor ($D_+ \propto a$).

Dark energy affects the growth of cosmological structure formation directly through the role of the expansion history on the  gravitational instability in an expanding background. This leads to the appearance of the background evolution $H(z)$ in the equation for the growing mode of density perturbations $D_{+}$ (Eq:{ \ref {eq:growingmode}}).  We use the growth rate $f(z)$ defined as 
\begin{equation} \label{eq:growthrate}
f(z) =  \frac{d {\rm ln} D_{+}}{d {\rm ln} a} 
\end{equation}
to quantify the growth rate of perturbations. The observationally measurable function $f(z)$ imprints the dynamics of dark energy and is sensitive to any departure from the $\Lambda CDM$ model.

 \begin{figure}[h]
 \begin{center}
  \begin{subfigure}{0.34\textwidth}
\includegraphics[height=4cm]{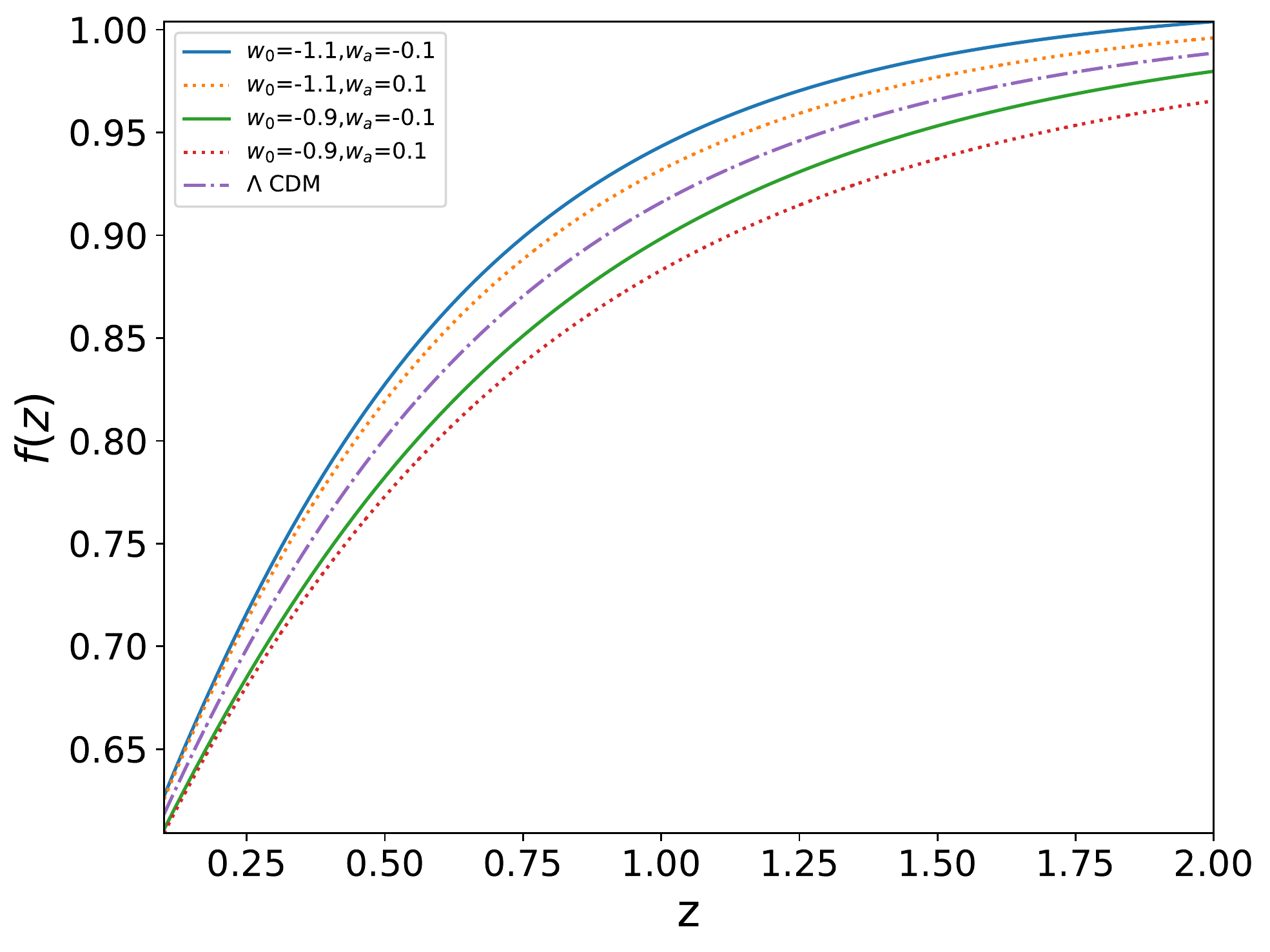}
\caption{CPL Model} \label{fig:1a}
\end{subfigure}%
  \hspace*{\fill} 
  \begin{subfigure}{0.34\textwidth}
  \includegraphics[height=4cm]{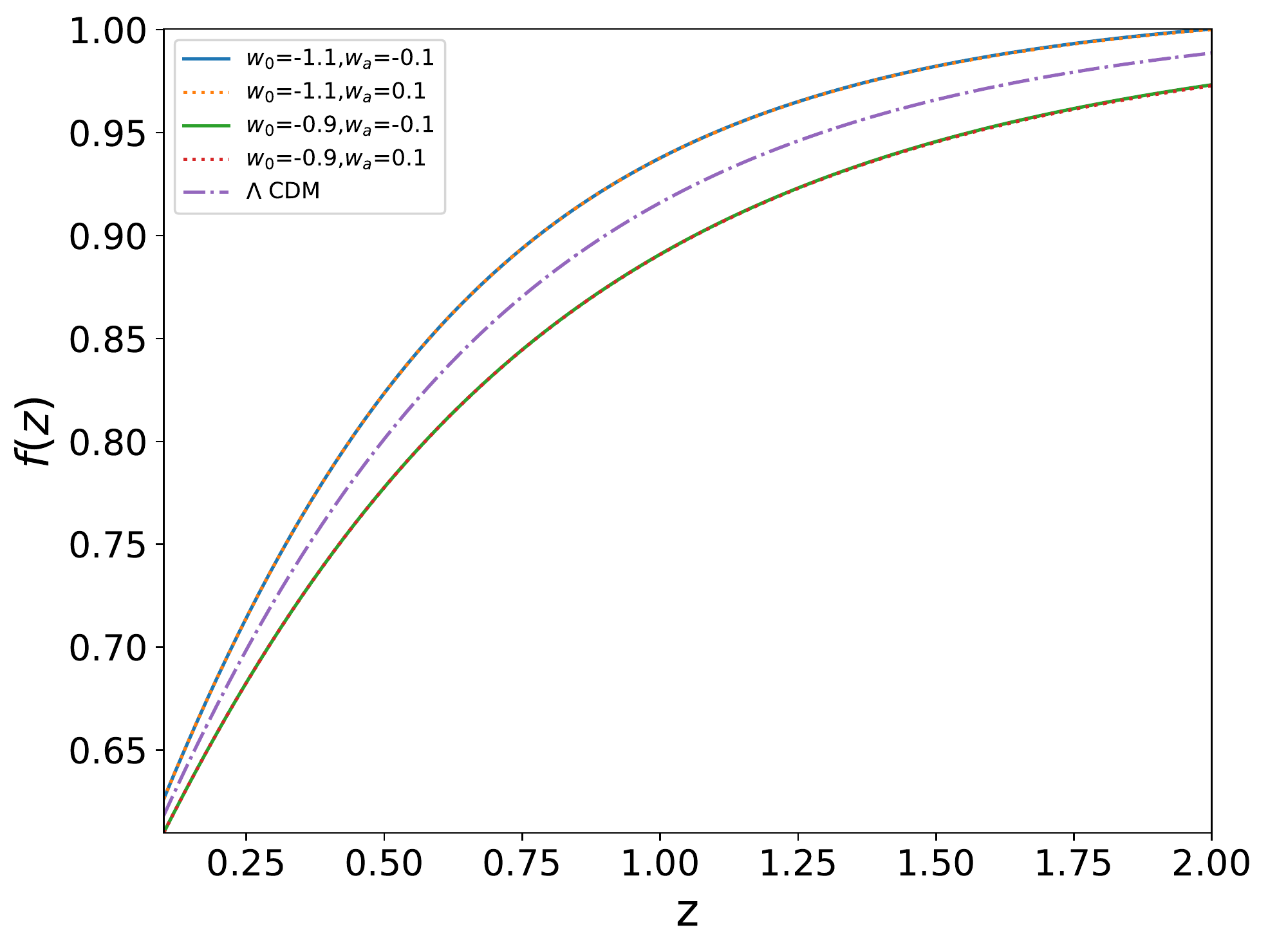}
\caption{7CPL Model} \label{fig:1b}
\end{subfigure}%
  \hspace*{\fill} 
\begin{subfigure}{0.34\textwidth}   
\includegraphics[height=4cm]{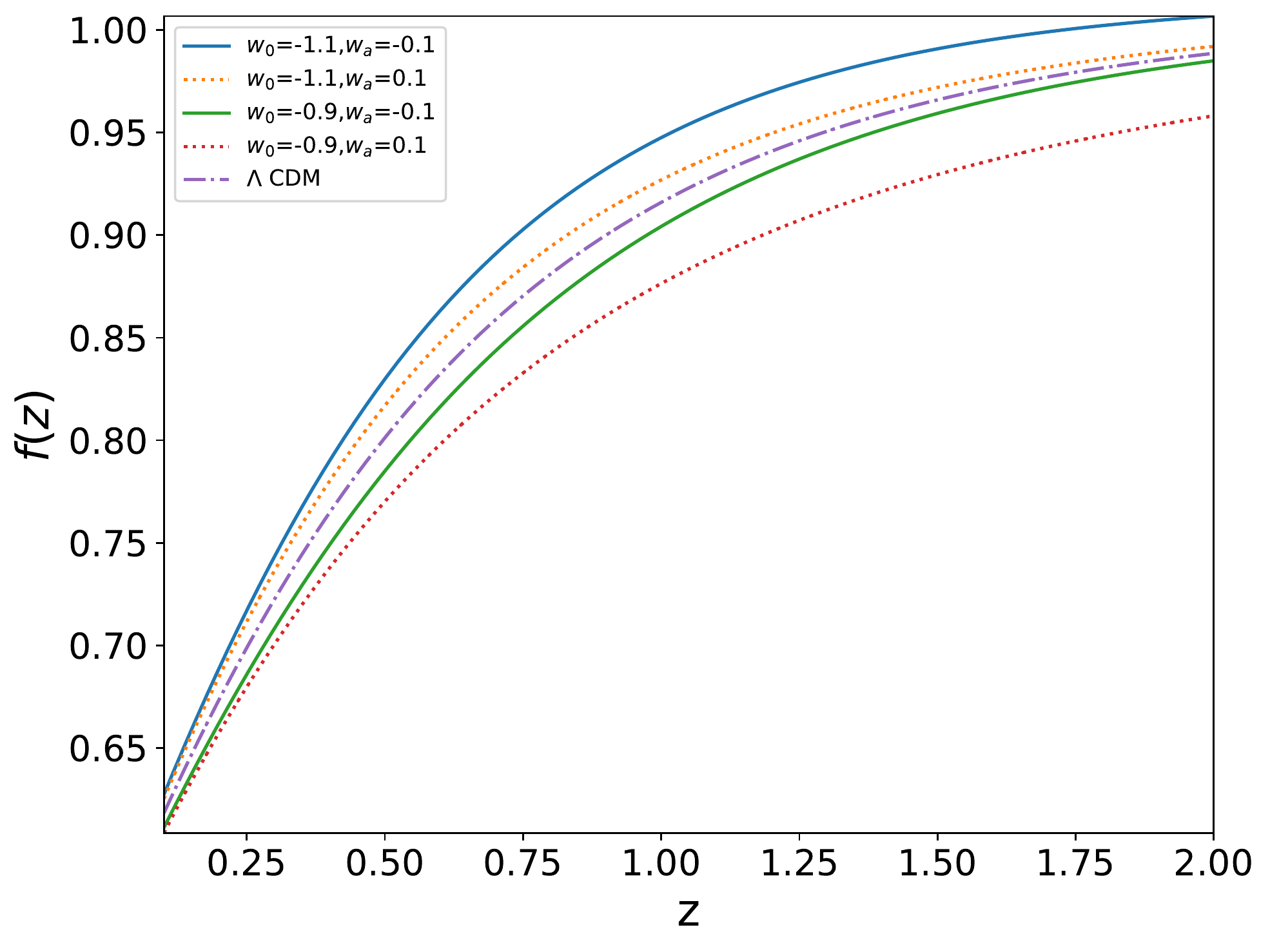}
\caption{BA Model} \label{fig:1c}
\end{subfigure}%
  \hspace*{\fill} 
\caption{ The growth rate of density fluctuations $f(z)$ for different dark energy EoS parametrizations. The $\Lambda CDM$ model is shown in each subfigure for comparison. }
\label{fig:growthrate}
\end{center}
\end{figure}
\begin{figure}[h]
 \begin{center}
  \begin{subfigure}{0.34\textwidth}
\includegraphics[height=4cm]{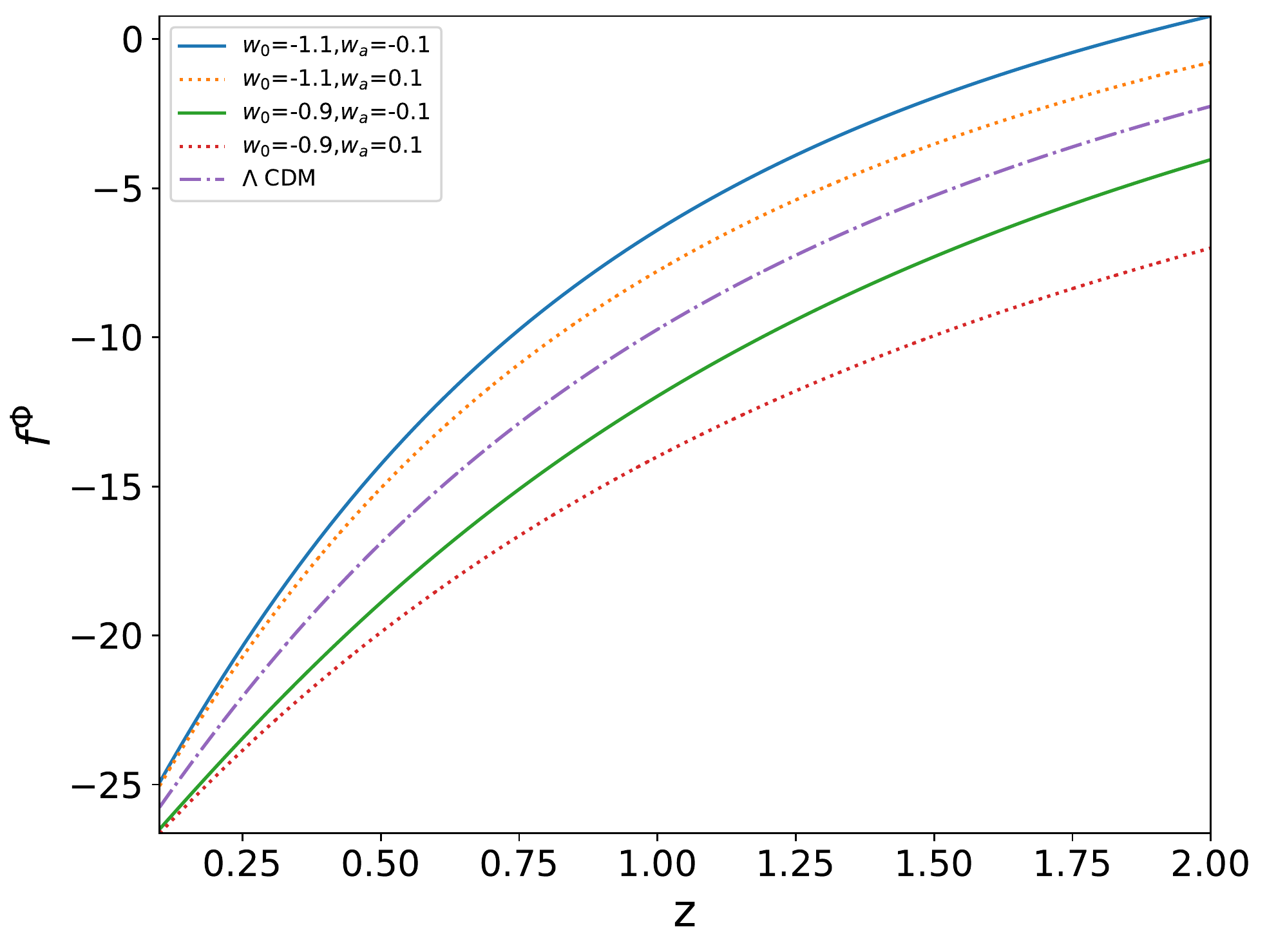}
\caption{CPL Model} \label{fig:isw1a}
\end{subfigure}%
  \hspace*{\fill} 
  \begin{subfigure}{0.34\textwidth}
  \includegraphics[height=4cm]{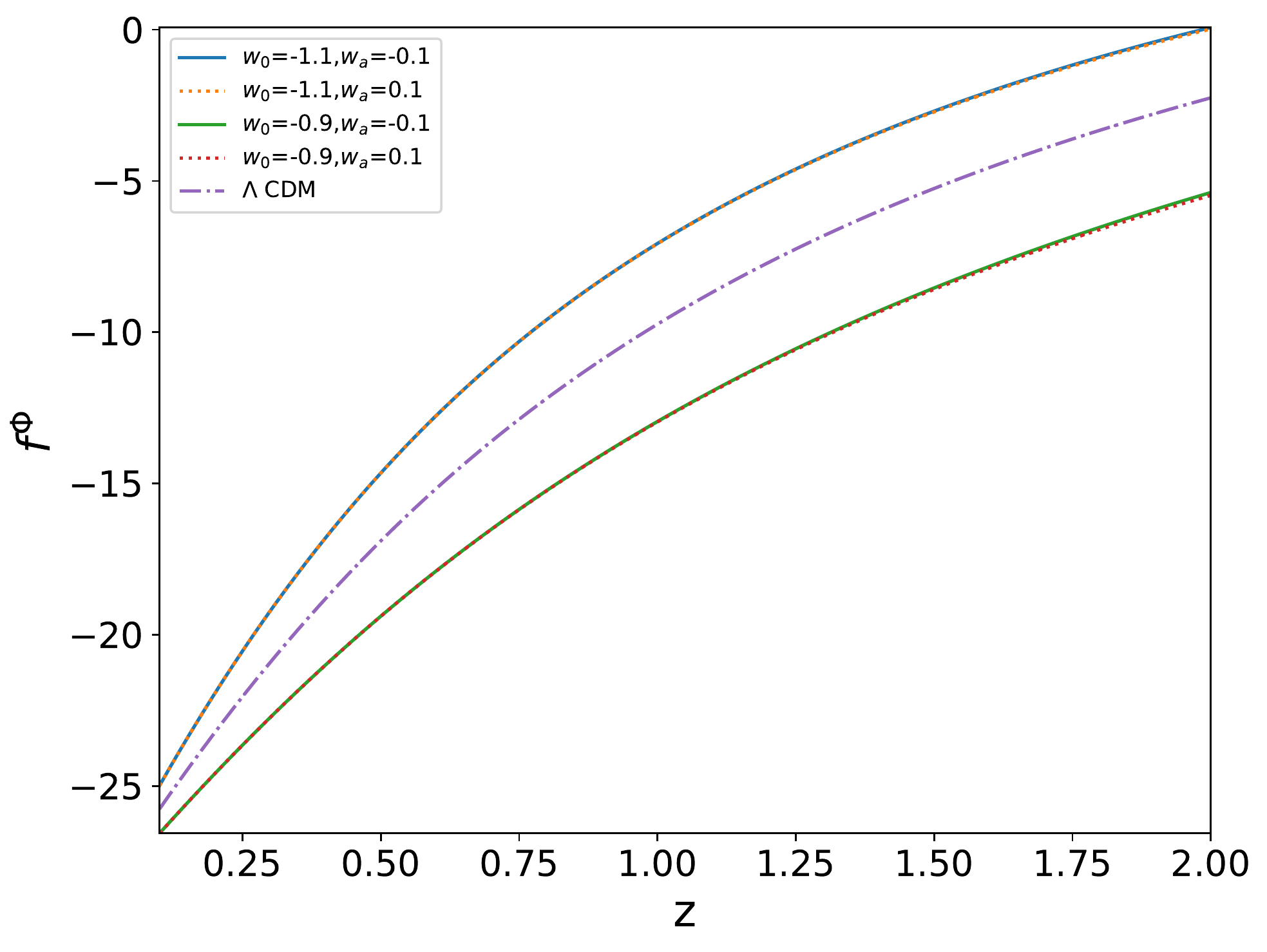}
\caption{7CPL Model} \label{fig:isw1b}
\end{subfigure}%
  \hspace*{\fill} 
\begin{subfigure}{0.34\textwidth}   
\includegraphics[height=4cm]{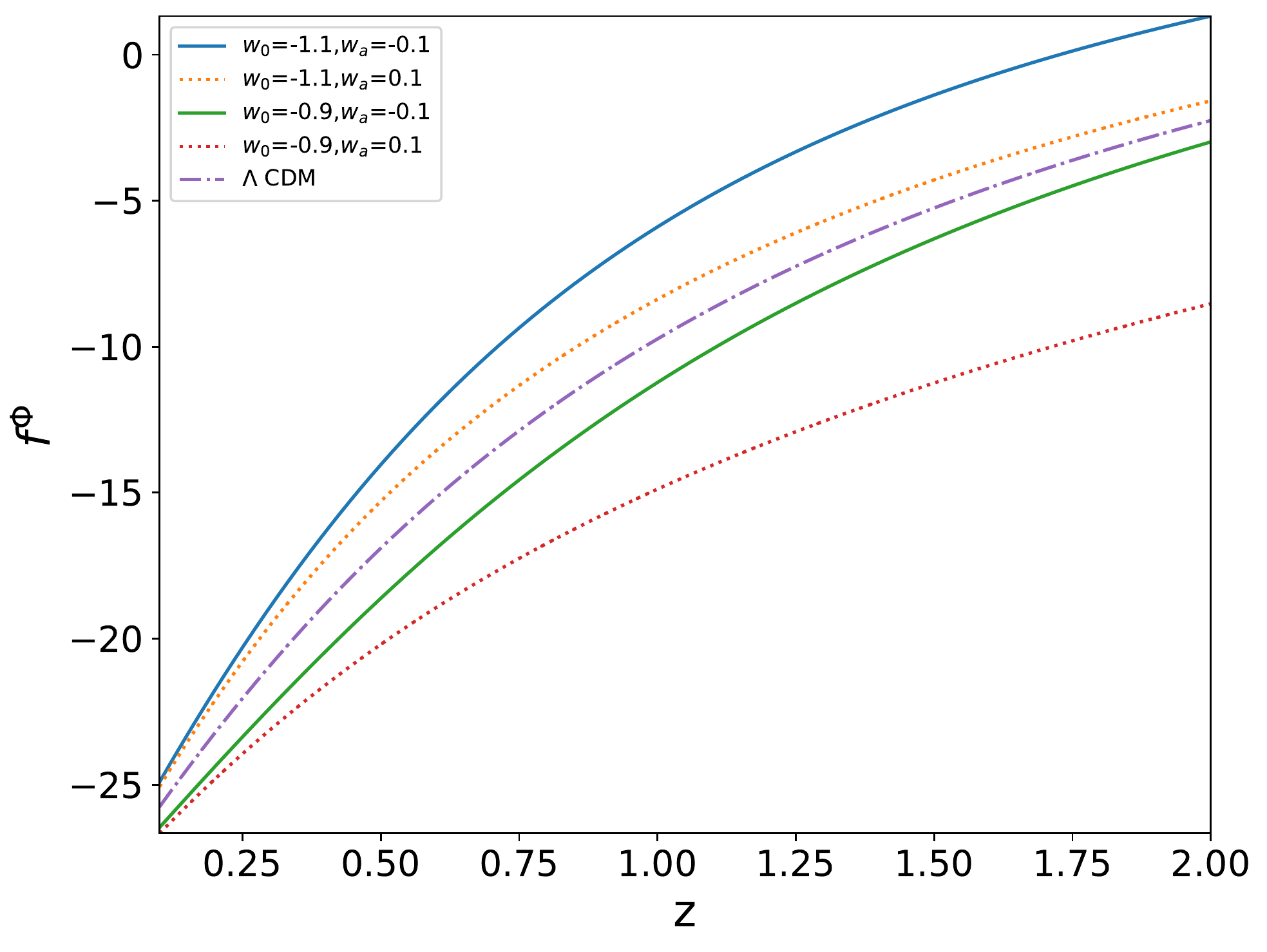}
\caption{BA Model} \label{fig:isw1c}
\end{subfigure}%
  \hspace*{\fill} 
\caption{The ISW parameter $f^{\Phi} = \frac{\dot \Phi}{\Phi}$ for  different dark energy EoS parametrizations. The $\Lambda CDM$ model is shown in each subfigure for comparison. }
\label{fig:growthrate-isw}
\end{center}
\end{figure}
Figure (\ref{fig:growthrate}), shows the linear growth rate of density perturbations  for CPL, 7CPL and BA parametrizations. At higher redshifts, the all models approach $ f \sim 1$ indicating that the growth of perturbation is dominated by non-relativistic matter. Further, for  7CPL model, there is no dependence of the parameter $w_a$.  Therefore, one does not expect to constrain the parameter $w_a$ for 7CPL model
using any observation related to matter clustering.

Dark energy also has an implicit observational effect. It causes a decay of the gravitational potential (the scalar perturbation in the Newtonian conformal gauge), when the
Universe evolves from the matter dominated to  the dark energy dominated era. This is known to generate a weak anisotropy in the CMB temperature fluctuation, through the Integrated
Sachs Wolfe (henceforth ISW) effect \citep{sachs_wolfe, Boughn_2005}. A curved  spatial geometry also similarly contributes to this anisotropy. However, the spatial curvature of our Universe is constrained to be zero from CMBR observations \cite{Komatsu_2009}
whereby the effect of spatial curvature can be ignored in the first approximation. The gravitational potential is expected to remain constant in a purely matter dominated cosmology. Thus, any late-time evolution of the gravitational potential is sensitive to the dark energy model \cite{scranton2003physical, PhysRevD.69.083524, Giannantonio_2008, Sarkar_2009}.
The ISW anisotropy is a line of sight integral \citep{hu2008lecture, k2004physics}
\be
\Delta T( {\bf \hat n}) ^{\rm ISW} = 2T \int_{\eta_{_{\rm LSS}}}^{\eta_0}  d\eta ~ \Phi'(r{\bf \hat n}, \eta) 
\ee
where $T$ is the present CMBR temperature, $\eta_{_{\rm LSS}}$ and $\eta_0$ are  the conformal times at the last scattering surface and the present epoch respectively and $\Phi' = d\Phi /d\eta$.
To quantify this effect we define a diagnostic for dark energy 
\be f^{\Phi} (z) =  \frac{d}{d \eta } {\rm ln} \Phi =  (f -1) \mathcal{H}  \ee
where ${\cal H} =  \frac{1}{a} da/d\eta$.  It is known that  $f \sim 1$ in pure matter dominated epoch and any departure from $f = 1$ indicates the action of dark energy. Noting that  the role of dark energy is 
 imprinted  in $f-1$, the function $f^{\Phi}(z)$ is a sensitive probe of dark energy. It quantifies the interplay of two time scales - the background expansion rate (contained in $H(z)$) and the growth rate of cosmological structure (contained in $f(z)$).

Figure(\ref{fig:growthrate-isw}) shows the redshift dependence of $f^{\Phi}(z)$ for different dark energy EoS parametrizations. The 7CPL model shows insensitivity to the parameter $w_a$, however both CPL and BA parametrizations show significant $ \sim 2-4 \%$ departure from the $\Lambda CDM$ behaviour. This may be crucial in  the improvement of detection sensitivities for ISW measurements \citep{Sarkar_2009}.

\end{document}